\documentclass[twocolumn]{aastex62}

\usepackage{color}
\usepackage{natbib}
\usepackage{amsmath}
\usepackage{comment}

\graphicspath{{./}{figures/}}

\received{\today}
\revised{\today}
\accepted{\today}
\submitjournal{ApJ}

\shorttitle{3D Boltzmann}
\shortauthors{Iwakami et al.}

\begin{document}

\title{SIMULATIONS OF THE EARLY POST-BOUNCE PHASE OF CORE-COLLAPSE SUPERNOVAE IN THREE-DIMENSIONAL SPACE WITH FULL BOLTZMANN NEUTRINO TRANSPORT}

\correspondingauthor{Wakana Iwakami}
\email{wakana@heap.phys.waseda.ac.jp}

\author{Wakana Iwakami}
\affiliation{Yukawa Institute for Theoretical Physics, Kyoto University, Oiwake-cho, Kitashirakawa, Sakyo-ku, Kyoto, 606-8502, Japan}
\affiliation{Advanced Research Institute for Science and Engineering, Waseda University, 3-4-1 Okubo, Shinjuku, Tokyo 169-8555, Japan}

\author{Hirotada Okawa}
\affiliation{Waseda Institute for Advanced Study, Waseda University, 1-6-1 Nishi-Waseda, Shinjuku-ku, Tokyo, 169-8050, Japan}

\author{Hiroki Nagakura}
\affiliation{Department of Astrophysical Sciences, Princeton University, Princeton, NJ 08544, USA}

\author{Akira Harada}
\affil{Institute for Cosmic Ray Research, University of Tokyo, 5-1-5 Kashiwanoha, Kashiwa, Chiba 277-8582, Japan}

\author{Shun Furusawa}
\affiliation{Department of Physics, Tokyo University of Science, Shinjuku, Tokyo,
162-8601, Japan}
\affiliation{Interdisciplinary Theoretical and Mathematical Sciences Program (iTHEMS), RIKEN, Wako, Saitama 351-0198, Japan}

\author{Kosuke Sumiyoshi}
\affiliation{Numazu College of Technology, Ooka 3600, Numazu, Shizuoka 410-8501, Japan}

\author{Hideo Matsufuru}
\affiliation{High Energy Accelerator Research Organization, 1-1 Oho, Tsukuba, Ibaraki 305-0801, Japan}

\author{Shoichi Yamada}
\affiliation{Advanced Research Institute for Science and Engineering, Waseda University, 3-4-1 Okubo, Shinjuku, Tokyo 169-8555, Japan}



\begin{abstract}
We report on the core-collapse supernova simulation we conducted for a 11.2$M_{\odot}$ progenitor model in three-dimensional space up to 20 ms after bounce, using a radiation hydrodynamics code with full Boltzmann neutrino transport.
We solve the six-dimensional Boltzmann equations for three neutrino species and the three-dimensional compressible Euler equations with Furusawa and Togashi’s nuclear equation of state.
We focus on the prompt convection at $\sim10$ ms after bounce and investigate how neutrinos are transported in the convective matter.
We apply a new analysis based on the eigenvalues and eigenvectors of the Eddington tensor and make a comparison between the Boltzmann transport results and the M1 closure approximation in the transition regime between the optically thick and thin limits.
We visualize the eigenvalues and eigenvectors using an ellipsoid, in which each principal axis is parallel to one of the eigenvectors and has a length proportional to the corresponding eigenvalue.
This approach enables us to understand the difference between the Eddington tensor derived directly from the Boltzmann simulation and the one given by the M1 prescription from a new perspective.
We find that the longest principal axis of the ellipsoid is almost always nearly parallel to the energy flux in the M1 closure approximation whereas in the Boltzmann simulation it becomes perpendicular in some transition regions, where the mean free path is $\sim0.1$ times the radius. In three spatial dimensions, the convective motions make it difficult to predict where this happens and to possibly improve the closure relation there.
\end{abstract}



\section{Introduction}

The explosion mechanism of core-collapse supernovae has been studied for a long time \citep[see][for reviews]{Janka2012, Burrows2013, Muller2016, Janka2016}.
Although it is not completely understood yet, accurate simulations in one spatial dimension (1D) under spherical symmetry have shown that the delayed explosion scenario by neutrino heating fails for most of progenitor models \citep[e.g.][]{Liebendorfer2001, Sumiyoshi2005}, whereas multi-dimensional simulations with some approximate neutrino transport have brought successful explosions \citep[e.g.][for recent papers]{Lentz2015, Pan2016, Takiwaki2016, Bruenn2016, Muller2017, Just2018, Vartanyan2019}.
However, it has not been clearly demonstrated whether the approximations for neutrino transport are really justified in the transition region between the optically thin and thick limits particularly in the multi-dimensional settings.
In fact, the turbulence energized by the neutrino-driven convection and the standing accretion shock instability form complex structures in density, electron fraction, and temperature in the transition region.
It is hence very interesting to see how the neutrinos are transported through such a chaotic environment.

Motivated by these arguments, we have developed the Boltzmann-radiation-hydrodynamics code for recent years: 
the basic framework of our code is given in \citet{Sumiyoshi2012}; 
the non-relativistic Boltzmann equation with the standard weak interactions is solved on some fixed matter distributions derived from supernova simulations \citep{Sumiyoshi2015}; 
the Boltzmann neutrino transport solver is extended so that it could treat the special relativistic effects to all orders of $v/c$, where $v$ and $c$ are the speeds of matter and light, respectively; 
the Boltzmann solver is coupled to a hydrodynamics solver with Newtonian self-gravity \citep{Nagakura2014}; 
a moving-mesh technique to follow the proper motion of proto-neutron star (PNS) is implemented in the general relativistic Boltzmann solver in the 3+1 formalism \citep{Nagakura2017}; 
very recently a new method to improve the accuracy of the momentum feedback from neutrino to matter is proposed by \citet{Nagakura2019}, and weak interactions for light nuclei are added to the code in \citet{Nagakura2019c}.

These codes have been employed in spatially two-dimensional (2D) simulations under axisymmetry to study the EOS dependence \citep{Nagakura2018} of and rotational effects \citep{Harada2019} on dynamics as well as possible early PNS kicks \citep{Nagakura2019a} and fast neutrino flavor conversions \citep{DelfanAzari2019,Nagakura2019b}.
In particular, \citet{Nagakura2018} and \citet{Harada2019} revealed non-axisymmetric features in the neutrino distribution function in momentum space and some differences in the Eddington tensors between the Boltzmann simulations and the M1 closure approximation. 
In this paper, we present for the first time the results of radiation hydrodynamic simulations in three spatial dimensions (3D) without any symmetry with our Boltzmann-radiation-hydrodynamics code.
We pay particular attentions to the neutrino angular distributions in the early post-bounce phase when the prompt convection grows.
We compare the Eddington tensor obtained in the Boltzmann simulations with those evaluated in the M1 closure approximation, employing a new analysis method.

This paper is organized as follows.
We explain the basic equations, weak interactions, and numerical setup used in this paper in Section \ref{sec:numerical}.
We first describe the 3D features in matter flows and radiation fields, comparing them with the 1D and 2D counterparts in Section \ref{sec:dynamical}.
We then conduct the new analysis with the eigenvalues and eigenvectors of the Eddington tensor in Section \ref{sec:analysis}.
We conclude this paper in Section \ref{sec:conclusion}.
Throughout the paper, the metric signature is $-\ +\ +\ +$, and we use the units with $c=G=h=1$ unless otherwise stated, where $G$ and $h$ are the gravitational constant and Planck's constant, respectively.

\section{Numerical Modeling \label{sec:numerical}}

\begin{figure*}[ht!]
\begin{center}
\includegraphics[width=\hsize]{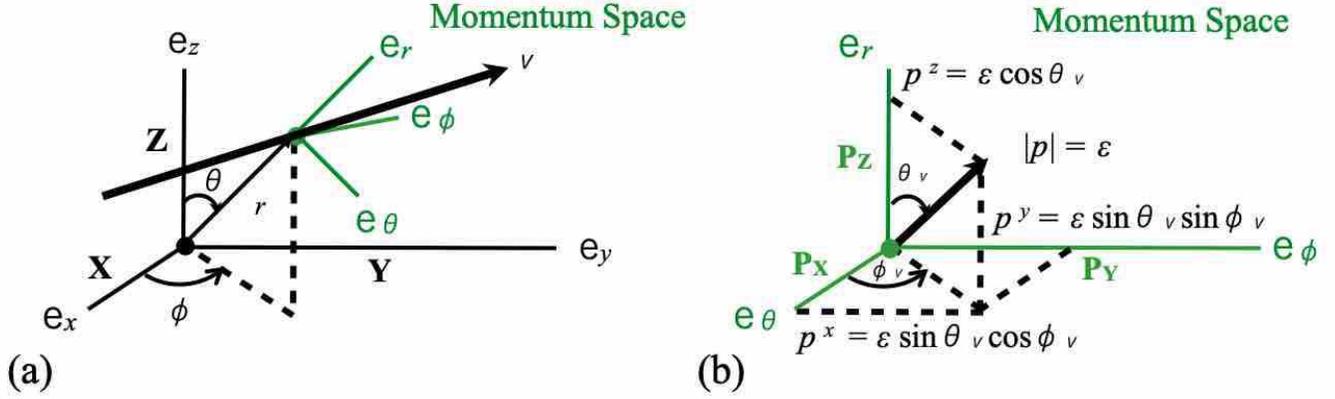}
\end{center}
\caption{The schematic pictures of the coordinate system in the phase space, where $r$, $\theta$, $\phi$, $\epsilon_\nu$, $\theta_\nu$, and $\phi_\nu$ are the radius, polar and azimuthal angles in space and the neutrino energy, and polar and azimuthal angles in momentum space, respectively.
The three orthogonal axes in space are labeled as $\mathrm{X}$, $\mathrm{Y}$, and $\mathrm{Z}$ (left panel), and those in momentum space are referred to as $\mathrm{P_X}$, $\mathrm{P_Y}$, and $\mathrm{P_Z}$ (right panel).
The directions of the three orthogonal axes $\mathrm{P_X}$, $\mathrm{P_Y}$, and $\mathrm{P_Z}$ in momentum space are chosen to agree with those of the basis vectors $\mathbf{e}_\theta$,  $\mathbf{e}_\phi$, and $\mathbf{e}_r$ in space, respectively.
The angle from the $\mathrm{P_Z}$ axis is defined as $\theta_\nu$, and the angle from the $\mathrm{P_X}$ axis on $\mathrm{P_X-P_Y}$ plane is denoted by $\phi_\nu$.
The distance from the origin in momentum space corresponds to the neutrino energy $\epsilon$.
\label{fig:coordinate}}
\end{figure*}

We use the Boltzmann-radiation-hydrodynamics code for core-collapse simulations \citep{Sumiyoshi2012,Nagakura2014}.
The coordinate system in phase space $(r, \theta, \phi, \epsilon, \theta_\nu, \phi_\nu)$ is shown in Figure \ref{fig:coordinate}, where $r$, $\theta$, $\phi$, $\epsilon$, $\theta_\nu$, and  $\phi_\nu$ denote radius, polar and azimuthal angles in space, neutrino energy, and polar and azimuthal angles in momentum space, respectively.
The special relativistic Boltzmann equation in the laboratory frame,
\begin{eqnarray}
    \frac{\partial f}{\partial t}
    +\frac{\mu_\nu}{r^2}\frac{\partial}{\partial r}(r^2 f)
    +\frac{\sqrt{1-\mu_\nu^2}\cos\phi_\nu}{r\sin\theta}\frac{\partial}{\partial\theta}(\sin\theta f)\nonumber\\
    +\frac{\sqrt{1-\mu_\nu^2}\sin\phi_\nu}{r\sin\theta}\frac{\partial f}{\partial\phi}
    +\frac{1}{r}\frac{\partial}{\partial \mu_\nu}[(1-\mu_\nu^2)f]\nonumber\\
    -\frac{\sqrt{1-\mu_\nu^2}}{r}\frac{\cos\theta}{\sin\theta}\frac{\partial}{\partial \phi_\nu}(\sin\phi_\nu f)=\left(\frac{\delta f}{\delta t}\right)_{\mathrm{col}},
    \label{eq:boltzmann}
\end{eqnarray}
is solved by the discrete ordinate $S_N$ method for three neutrino species: $\nu_e$, $\bar{\nu}_e$, and $\nu_x$; $f$ and $\mu_\nu\  (=\cos\theta_\nu)$ are the neutrino distribution function in phase space and cosine of the polar angle in momentum space, respectively.
Using the finite volume method, we also solve the Newtonian compressible hydrodynamics equations along with the time evolution equation of electron number density:
\begin{eqnarray}
\partial_t \mathbf{Q}+\partial_j \mathbf{U}^j=
\mathbf{W}_h + \mathbf{W}_i, 
\end{eqnarray}
\begin{eqnarray}
\mathbf{Q}=\sqrt{g}\left(\begin{array}{c}
     \rho \\
     \rho v_r \\
     \rho v_\theta \\
     \rho v_\phi \\
     e + \frac{1}{2}\rho v^2 \\
     \rho Y_e
\end{array}\right),
\end{eqnarray}
\begin{eqnarray}
\mathbf{U}^j=\sqrt{g}\left(\begin{array}{c}
     \rho v^j\\
     \rho v_r v^j + P \delta^j_r\\
     \rho v_\theta v^j  + P \delta^j_\theta\\
     \rho v_\phi v^j + P \delta^j_\phi\\
     \left(e + P + \frac{1}{2}\rho v^2\right)v^j \\
     \rho Y_e v^j
\end{array}\right),
\end{eqnarray}
\begin{eqnarray}
\mathbf{W}_h=\sqrt{g}\left(\begin{array}{c}
     0 \\
     -\rho\partial_r \Psi 
     + \rho r (v^\theta)^2
     + \rho r \sin^2\theta(v^\phi)^2
     +\frac{2P}{r}\\
     -\rho r^2\partial_\theta \Psi
     +\rho\sin\theta \cos\theta (v^\phi)^2
     +\frac{P\cos\theta}{\sin\theta}\\
     -\rho \partial_\phi \Psi\\
     -\rho v^j \partial_j \Psi\\
     0
\end{array}\right),
\end{eqnarray}
\begin{eqnarray}
\mathbf{W}_i=\sqrt{g}\left(\begin{array}{c}
     0 \\
     -G_r \\
     -G_\theta \\
     -G_\phi \\
     G_t \\
     -\Gamma
\end{array}\right),
    \label{eq:hydro4}
\end{eqnarray}
where $\rho$, $v^j$, $e$, $Y_e$, $P$, $\Psi$, $\delta^j_i$, and $g\ (=r^2\sin\theta)$ are the mass density, fluid velocity, internal energy density, electron fraction, pressure, Newtonian gravitational potential, Kronecker's delta, and volume factor in the spherical coordinates, respectively.
Here the index $j$ runs over the values 1, 2, 3, which correspond to the three components of spatial coordinates.
The numerical flux is determined by the HLL scheme with the piecewise-parabolic interpolation, and the time integration is done with the third order Runge-Kutta method.
Unlike in the papers by \citet{Nagakura2018} and \citet{Harada2019}, we ignore the term of the coordinate acceleration $\mathbf{W}_a$ given in \citet{Nagakura2017} and assume the monopole gravity: the gradient of $\Psi$ is obtained as follows
\begin{eqnarray}
\partial_j \Psi =\left(\frac{M(r)}{r^2}, 0, 0\right),\\
M(r) = \int_0^r  dr \int  d\Omega \ r^2 \rho (r, \theta, \phi) ,
\end{eqnarray}
where $d\Omega\ (=\sin\theta d\theta d\phi)$ is the differential solid angle in the spherical coordinates.
Equation (\ref{eq:hydro4}) presents the feedback from neutrino to hydrodynamics,
\begin{eqnarray}
G^\alpha \equiv \sum_s G^\alpha_s,\\
G^\alpha_s \equiv \int p^\alpha_s \left(\frac{\delta f}{\delta \lambda}\right)_{\mathrm{col}(s)} dV_p,
\label{eq:interaction1}
\end{eqnarray}
\begin{eqnarray}
\Gamma \equiv \Gamma_{\nu_e}-\Gamma_{\bar{\nu}_e},\\
\Gamma_s \equiv \int \left(\frac{\delta f}{\delta \lambda} \right)_{\mathrm{col}(s)} dV_p,
\label{eq:interaction3}
\end{eqnarray}
where $p^\alpha_s$, $\lambda$, and $dV_p$ denote the four momentum of neutrino of species $s$, affine parameter, invariant volume in momentum space, respectively.
The index $\alpha$ runs over the four values 0, 1, 2, 3, corresponding to the time and space components.
The collision term $(\delta f/\delta \lambda)_{\mathrm{col}}$ is related to $(\delta f/\delta t)_{\mathrm{col}}$ in the laboratory frame and $(\delta f/\delta \tau)_{\mathrm{col}}$ in the fluid-rest frame as follows,
\begin{eqnarray}
\left(\frac{\delta f}{\delta \lambda}\right)_{\mathrm{col}}
=\epsilon \left(\frac{\delta f}{\delta  t}\right)_{\mathrm{col}}
=\epsilon_\mathrm{FR}\left(\frac{\delta f}{\delta \tau} \right)_{\mathrm{col}},
\label{eq:col}
\end{eqnarray}
where $\tau$ and $\epsilon_\mathrm{FR}$ are the proper time of fluid element and the neutrino energy measured in the fluid-rest frame, respectively.

We consider the following neutrino reactions;

\noindent the absorption/emissions:
\begin{eqnarray}
(\mathrm{ecp})&:&\ e^- + p \longleftrightarrow \nu_e + n,\\
(\mathrm{aecp})&:&\ e^+ + n \longleftrightarrow \bar{\nu}_e + p,\\
(\mathrm{eca})&:&\ e^- + A \longleftrightarrow \nu_e + A',
\end{eqnarray}
the scatterings:
\begin{eqnarray}
(\mathrm{nsc})&:&\ \nu_s + N \longleftrightarrow \nu_s + N,\\
(\mathrm{csc})&:&\ \nu_s + A \longleftrightarrow \nu_s + A,\\
(\mathrm{esc})&:&\ \nu_s + e \longleftrightarrow \nu_s + e,
\end{eqnarray}
and the pair processes:
\begin{eqnarray}
(\mathrm{pap})&:&\ e^-+e^{+} \longleftrightarrow \nu_s +\bar{\nu}_s,\\
(\mathrm{nbr})&:&\ N+N \longleftrightarrow N+N +\nu_s +\bar{\nu}_s,
\end{eqnarray}
where the subscript $s$ again denotes the neutrino species.
The detailed expressions of the collision terms are given in \citet{Sumiyoshi2012}.
Most of the reaction rates are taken from \citet{Bruenn1985}.
The weak interactions of electron neutrino  (ecl) with light nuclei:
\begin{eqnarray}
(\mathrm{elpp})&:&\ \nu_e+ {}^2\mathrm{H} \longleftrightarrow e^- + p + p, \\
(\mathrm{el2h})&:&\ \nu_e+n+n \longleftrightarrow e^- + {}^2\mathrm{H}, \\
(\mathrm{el3h})&:&\ \nu_e+{}^3\mathrm{H}  \longleftrightarrow e^- + {}^3\mathrm{He},
\end{eqnarray}
and those of anti-electron neutrino (aecl):
\begin{eqnarray}
(\mathrm{ponn})&:&\ \bar{\nu}_e+ {}^2\mathrm{H} \longleftrightarrow e^+ + n + n, \\
(\mathrm{po2h})&:&\ \bar{\nu}_e+p+p \longleftrightarrow e^+ + {}^2\mathrm{H}, \\
(\mathrm{po3h})&:&\ \bar{\nu}_e+{}^3\mathrm{He}  \longleftrightarrow e^+ + {}^3\mathrm{H}, 
\end{eqnarray}
are also taken into account.
We use the multi-compositional equation of state (EOS) at sub-nuclear densities \citep{Furusawa2017} based on the supra-nuclear density EOS calculated with the variational method \citep{Togashi2013} and the new tables for electron capture by heavy and light nuclei and positron capture by light nuclei \citep{Nagakura2019}.

\begin{figure}[t!]
\begin{center}
\includegraphics[width=5.5 cm]{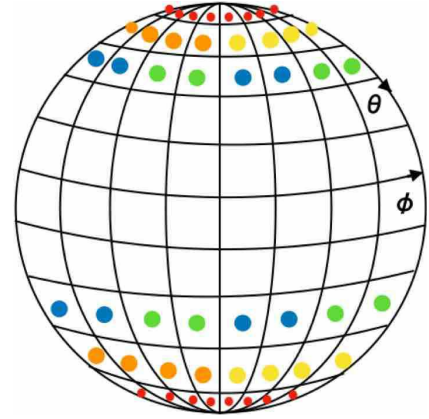}
\end{center}
\caption{The schematic image of a numerical spatial grid in a spherical coordinate system. The conserved variables are averaged over the same colored cells in the vicinity of the polar axis for the numerical stability.
\label{fig:grid}}
\end{figure}

In this study, we employ the 11.2$M_\odot$ progenitor model in \citet{Woosley2002}.
The spatial domain of $0\le r\le5000$ km is divided into 384 radial grid cells in 1D simulations.
The number is reduced to 256 and the spatial domain covers only $0\le r \le 200$ km in multi-dimensional simulations.
The entire solid angle in space is divided into 48 grid cells in $\theta$ for 2D and 48$\times$96 grid cells in $\theta \times \phi$ for 3D simulations.
The neutrino energy in the range of $0 \le \epsilon \le$ 300 MeV is divided into 16 cells. 
The entire solid angle in momentum space is divided into 6 cells in $\theta_\nu$ for 2D and $6\times6$ cells in $\theta_\nu\times\phi_\nu$ for 3D simulations.
More detailed discussions on the numerical resolution are provided in Appendix \ref{sec:resolution}.

\begin{figure*}[ht!]
\begin{center}
\includegraphics[width=\hsize]{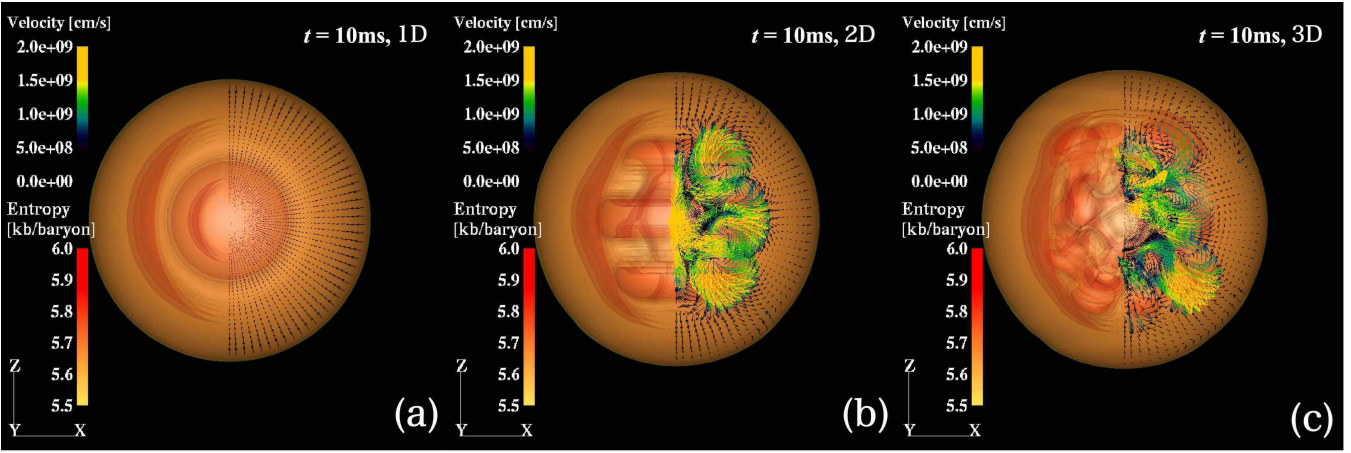}
\end{center}
\caption{The iso-surfaces of entropy with the fluid velocity vectors in (a) 1D, (b) 2D, and (c) 3D simulations at $t=10$ ms. 
The iso-surfaces are cut away between $\phi=225^\circ$ and $360^\circ$.
The vectors are superimposed only on the meridian plane in the right half part.
The shock wave is located at $r=70$ km, corresponding to the surfaces of orange spheres. 
\label{fig:Hydro_merid}}
\end{figure*}

The procedure of our 3D simulations is as follows:
\begin{enumerate}
\item the corresponding 1D simulation is done from the onset of the core-collapse to a shock stagnation,
\item the time dependent boundary data are extracted from the 1D results for the use in multi-dimensional simulations,
\item the multi-dimensional simulation is started with an introduction of 1\% random velocity perturbations on the spherically symmetric flow when a negative entropy gradient emerges for the first time,
\item the values of conserved variables at the outer boundary are obtained at every time step by linearly interpolating the boundary data obtained in step~2.
\end{enumerate}
Furthermore, two kinds of coarse graining are implemented to relax the Courant-Friedrichs-Lewy (CFL) condition for numerical stability around the singular points in spherical coordinates: near the coordinate origin ($0 < r < 8$ km), the physical quantities are averaged over the solid angle; in the vicinity of the polar axis, the conserved variables are averaged over every eight, four, and two cells in the $\phi$ direction for the first, second, and third $\theta$ cells from the polar axis, respectively (Fig.~\ref{fig:grid}).
Note that the moving mesh technique is not activated in this study.

\section{FEATURES OF FLOW AND RADIATION FIELDS \label{sec:dynamical}}

\begin{figure*}[ht!]
\begin{center}
\includegraphics[width=\hsize]{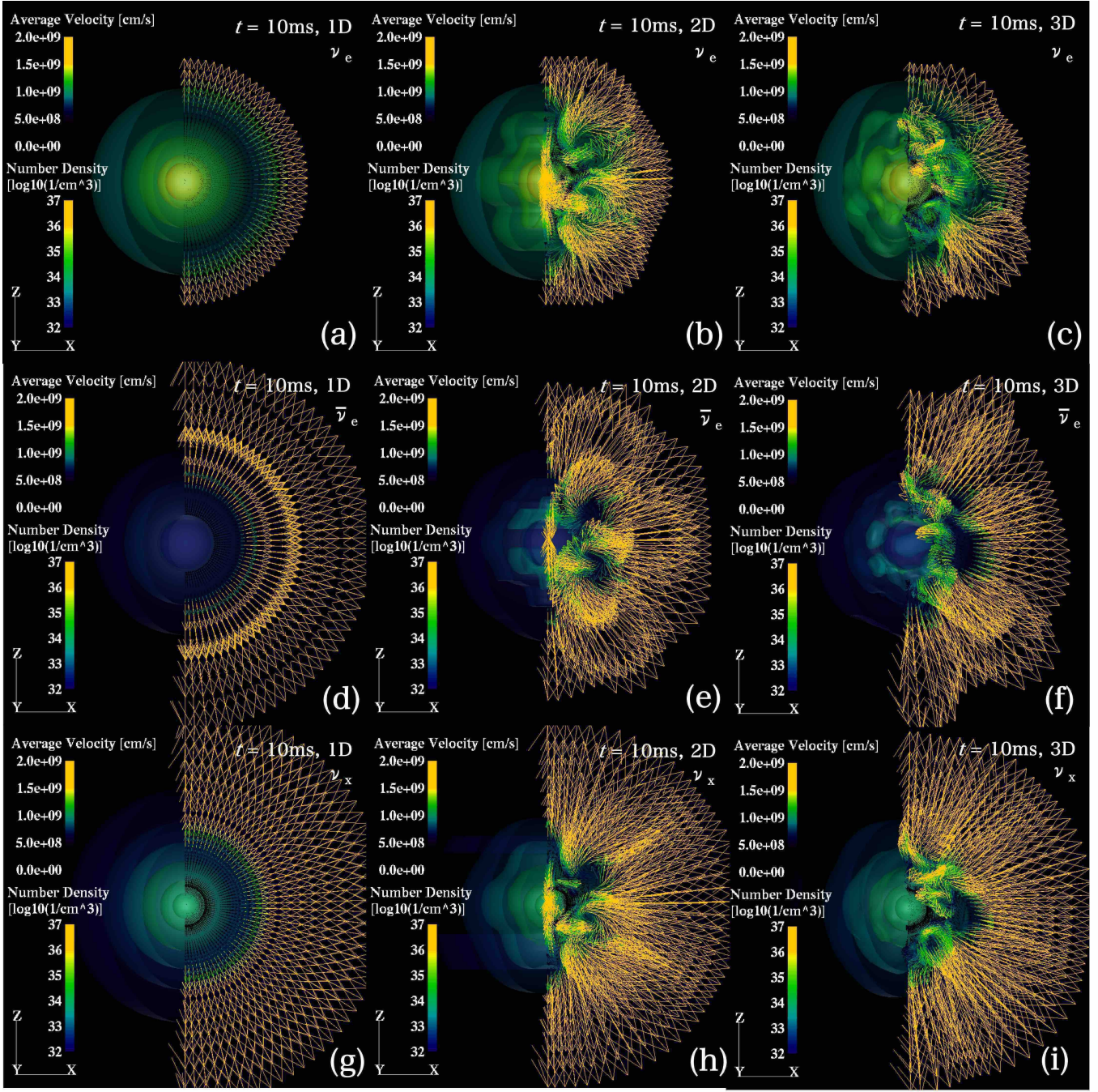}
\end{center}
\caption{The iso-surfaces of neutrino number density with the average velocity vectors for $\nu_e$ (1st rows), $\bar{\nu}_e$ (2nd row), and $\nu_x$ (3rd row) in 1D (left panels), 2D (middle panels) and 3D (right panels) simulations at $t=10$ ms. 
The iso-surfaces are cut away between $\phi=225^\circ$ and $360^\circ$. 
The vectors are superimposed only on the meridian planes in the right half part.
The vectors of the average velocity for $\bar{\nu}_e$ are not shown in $r<20$ km.
The spatial size is almost same as Fig.~\ref{fig:Hydro_merid}.
\label{fig:Neutrino_merid}}
\end{figure*}

\begin{figure*}[ht!]
\begin{center}
\includegraphics[width=\hsize]{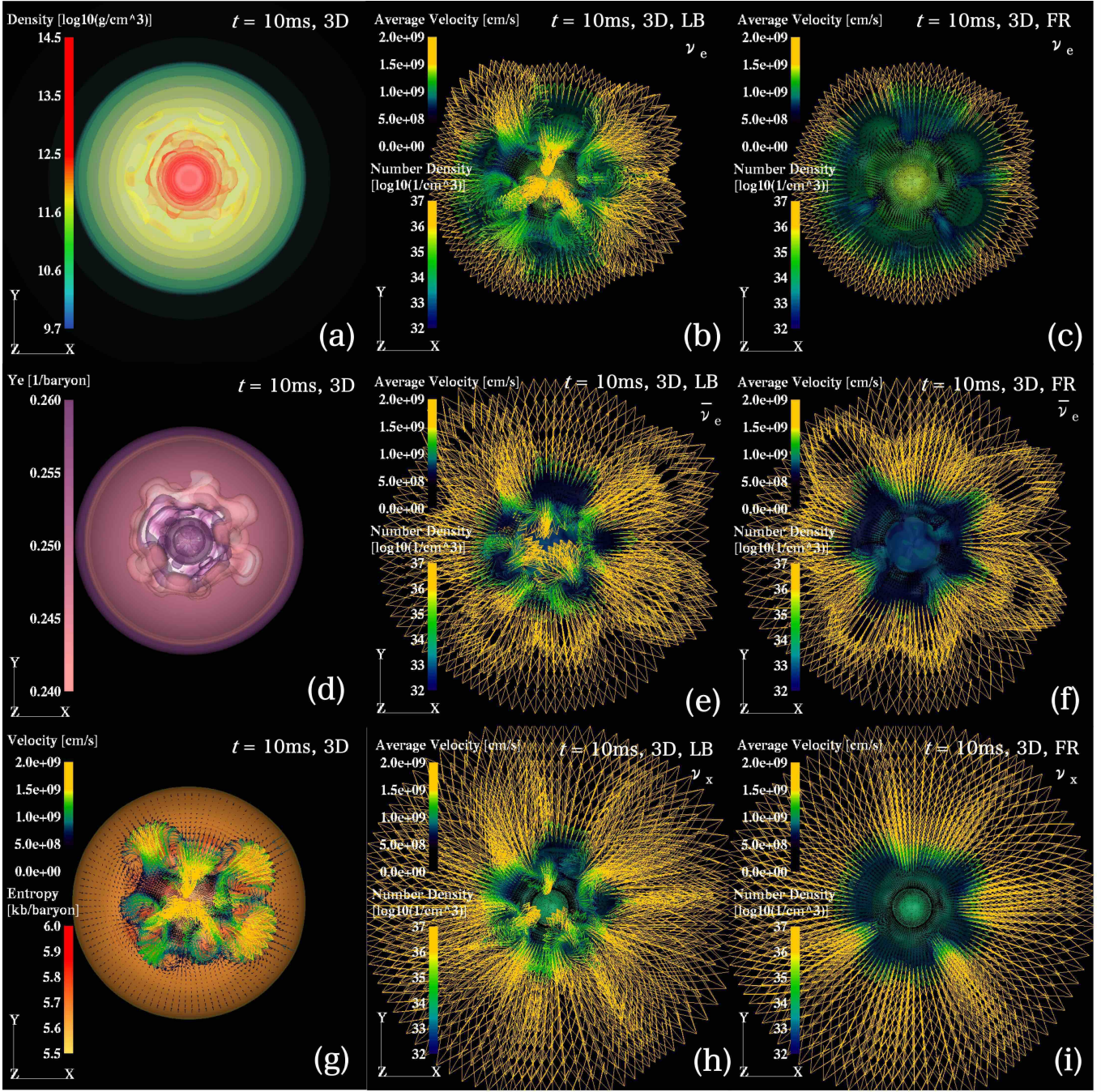}
\end{center}
\caption{
The isosurfaces of density, electron fraction, entropy with the fluid velocity vectors, are shown in the left column.
The isosurfaces of neutrino number densities with the average velocity vectors for $\nu_e$ (1st row),  $\bar{\nu}_e$ (2nd row), and $\nu_x$ (3rd row) measured in the laboratory frame (LB) and fluid-rest frame (FR) are also shown in the middle and right columns, respectively. 
The isosurfaces are cut away above the equatorial plane.
The vectors are superimposed on the cut planes.
The vectors of the average velocity for $\bar{\nu}_e$ are not shown in $r<20$ km.
\label{fig:Hydro_Neutrino_equat}}
\end{figure*}

\begin{figure*}[ht!]
\begin{center}
\includegraphics[width=\hsize]{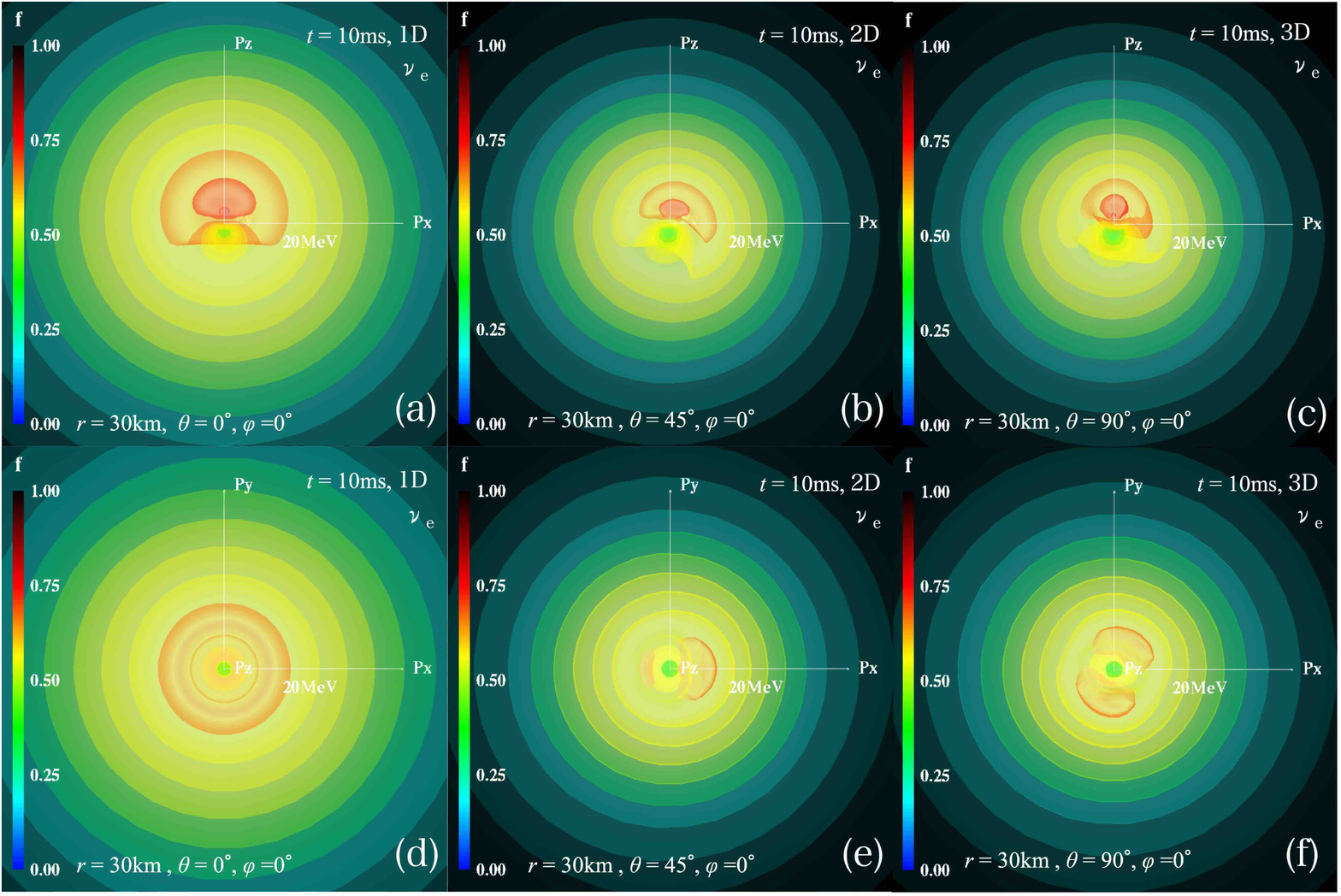}
\end{center}
\caption{The momentum space distribution of $f$ for $\nu_e$ at $r=30$ km, which is in the diffusion regime.
The distance from the origin corresponds to the neutrino energy $\epsilon$, and the length of the axes is 20 MeV.
The neutrino distribution function is axisymmetric around the $\mathrm{P_Z}$ axis in 1D, symmetric with respect to the $\mathrm{P_Z}-\mathrm{P_X}$ plane in 2D whereas it does not have any symmetry in 3D.
\label{fig:NeutrinoDistribution}}
\end{figure*}

We present some noteworthy features in matter flows and neutrino radiation fields in 1D, 2D, and 3D simulations.
Figures \ref{fig:Hydro_merid} and \ref{fig:Neutrino_merid} show, respectively, the isosurfaces of entropy with the fluid velocity vectors and those of neutrino number density with the average velocity vectors for $\nu_e$, $\bar{\nu}_e$, and $\nu_x$ (1st, 2nd, and 3rd rows) at $t=10$ ms.
The average velocity of neutrino $v^i_\nu$ is defined as
\begin{equation}
v^i_\nu=\frac{\mathcal{F}^i}{\mathcal{N}}.
\label{eq:vi}
\end{equation}
The neutrino number density $\mathcal{N}$ and number flux $\mathcal{F}^i$ are given as
\begin{eqnarray}
\mathcal{N} = \int \epsilon^2d\epsilon \int d\Omega_\nu \ f (\epsilon, \Omega_\nu), \label{eq:N}\\
\mathcal{F}^i = \int \epsilon^2d\epsilon \int d\Omega_\nu \ f (\epsilon, \Omega_\nu) \  n^i, \label{eq:Fi}
\end{eqnarray}
where $n^i$ and $d\Omega_\nu(=\sin\theta_\nu d\theta_\nu d\phi_\nu)$ are the unit vector in space and the differential solid angle in the spherical coordinates, respectively.
Here the integrations in Eqs.~(\ref{eq:N}) and (\ref{eq:Fi}) are done in the laboratory frame.
The isosurfaces are cut away between $\phi=225^\circ$ and $360^\circ$.
The vectors are superimposed only on the meridian plane in the right half part.
For better visibility, the vectors of the average velocity for $\bar{\nu}_e$ are not shown in $r < 20$ km.
The shock wave is located at $r=70$ km, corresponding to the surfaces of orange spheres. 
Unlike the results in 1D (Fig.~\ref{fig:Hydro_merid} (a)), the prompt convection grows inside the shock wave in the multi-dimensional computations.
Three vortex rings grow in 2D (Fig.~\ref{fig:Hydro_merid} (b)),
whereas multiple round convective vorticies are formed in 3D (Fig.~\ref{fig:Hydro_merid} (c)).
These convective flows affect the transport of $\nu_e$, $\bar{\nu}_e$, and $\nu_x$.
Multi-dimensional structures are also developed in the number densities of neutrinos both in 2D (Fig.~\ref{fig:Neutrino_merid} (b), (e), and (h)) and 3D (Fig.~\ref{fig:Neutrino_merid} (c), (f), and (i)).
In the central convective region, neutrinos move in various directions with matter.
In the outer region, they propagate outward both in 2D (Fig.~\ref{fig:Neutrino_merid} (b), (e), and (h)) and 3D (Fig.~\ref{fig:Neutrino_merid} (c), (f), and (i)) anisotropically.

Figure~\ref{fig:Hydro_Neutrino_equat} shows the isosurfaces of the density, electron fraction, entropy with the fluid velocity vectors, and the number density with the average velocity vectors for $\nu_e$, $\bar{\nu}_e$, and $\nu_x$ on the equatorial plane for 3D.
The iso-surfaces are cut away above the equatorial plane.
The vectors are superimposed on the cut planes.
The integrations in Eqs.~(\ref{eq:N}) and (\ref{eq:Fi}) are done both in the laboratory (LB) and fluid-rest (FR) frames.
The random perturbation added initially by hand grows exponentially in the region of the negative entropy gradient, and complex structures in density, electron fraction, entropy, and velocity develop as a result on the equatorial plane in the 3D simulation (Fig.~\ref{fig:Hydro_Neutrino_equat}~(a), (d), and (g)).
The spatial distributions of number density and average velocity for $\nu_e$, $\bar{\nu}_e$, and $\nu_x$ are also affected by them (Fig.~\ref{fig:Hydro_Neutrino_equat}~(b), (e), and (h)).
Although the radiation fields evolve differently among three species of neutrinos due to their interactions with matter in different ways,
they have a common feature.
The average neutrino velocity measured in the laboratory frame agrees roughly with the matter velocity in the central convective region  (Fig.~\ref{fig:Hydro_Neutrino_equat} (b), (e), (g), and (h)) whereas the average neutrino velocity in the fluid-rest frame is quite small in the same region (Fig.~\ref{fig:Hydro_Neutrino_equat} (c), (f), and (i)). 
This means that neutrinos are tightly coupled to matter in this optically thick region.
The two-energy grid approach enables us to capture this effect to the full order of $v/c$ \citep{Nagakura2014}. 
On the other hand, neutrinos begin to propagate outward freely in the outer region.
Our code can capture the transition between the optically thick and thin limits.

\begin{figure*}[ht!]
\begin{center}
\includegraphics[width=\hsize]{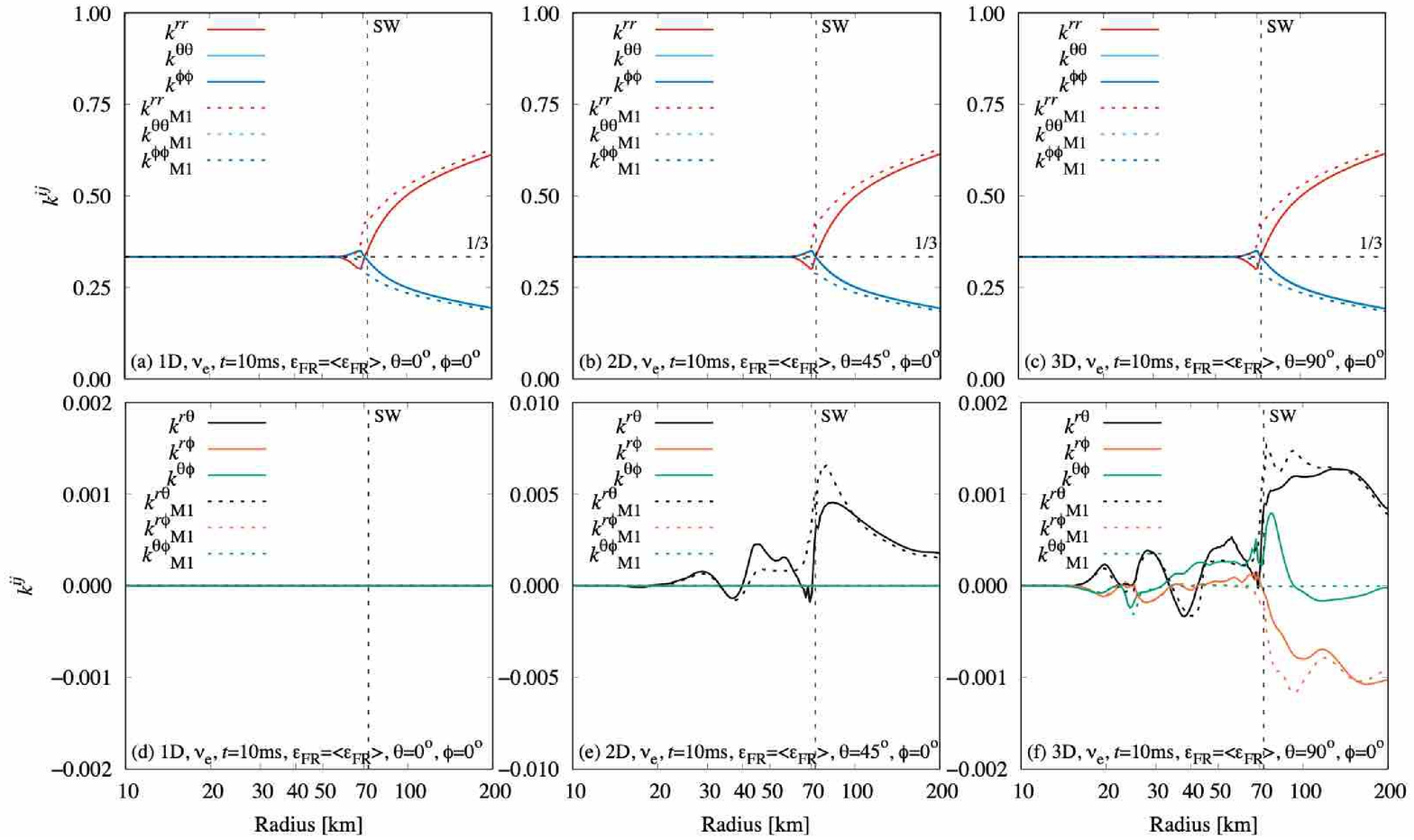}
\end{center}
\caption{The typical radial distributions of diagonal (upper panels) and off-diagonal (lower panels) components of Eddington tensors $k^{ij}$ (solid lines) and $k^{ij}_{M1}$ (dashed lines) at $\epsilon_\mathrm{FR}=\langle\epsilon_\mathrm{FR}\rangle$ for $\nu_e$ at $t=10$ ms in 1D (left panels), 2D (middle panels), and 3D (right panels) simulations, where $k^{ij}$, $k^{ij}_{M1}$, $\epsilon_\mathrm{FR}$, and $\langle\epsilon_\mathrm{FR}\rangle$ are the Eddington tensors calculated in the Boltzmann simulation and the M1 closure approximation, the neutrino energy observed in the fluid rest frame, and its averaged energy, respectively.
\label{fig:Eddington}}
\end{figure*}

Figure \ref{fig:NeutrinoDistribution} shows the isosurfaces of neutrino distribution function $f$ at $r=30$ km, which is in the diffusion regime.
The directions of the three orthogonal axes $\mathrm{P_X}$, $\mathrm{P_Y}$, and $\mathrm{P_Z}$ in momentum space are chosen to agree with those of the $\theta$-, $\phi$-, and $r$-axes in space, respectively.
The angle from the $\mathrm{P_Z}$ axis is defined as $\theta_\nu$, and the angle from the $\mathrm{P_X}$ axis on the $\mathrm{P_X-P_Y}$ plane is denoted by $\phi_\nu$ (See Fig.~\ref{fig:coordinate}).
The distance from the origin corresponds to the neutrino energy $\epsilon$, and the length of the axes is 20 MeV in these plots. 
The neutrino distribution function is axisymmetric around the $\mathrm{P_Z}$ axis in 1D (Fig.~\ref{fig:NeutrinoDistribution} (a) and (d)), symmetric with respect to the $\mathrm{P_Z-P_X}$ plane in 2D (Fig.~\ref{fig:NeutrinoDistribution} (b) and (e)) whereas it does not have any symmetry in 3D (Fig.~\ref{fig:NeutrinoDistribution} (c) and (f)).

The Eddington tensor is defined as the pressure tensor divided by the energy density.
In the Boltzmann neutrino radiation transport simulations, we can directly calculate it from the distribution function (see Appendix \ref{sec:eddington}).
The moment method is one of the common approximations for neutrino transport.
In the M1 closure approximation, in particular, the evolution equations of the energy density $E$ and flux $F^i$ are solved with
an artificial closure relation for the pressure tensor $P^{ij}$ \citep[e.g.][]{Thorne1981, Shibata2011}.
In the following paragraph, we compare the Eddington tensor $k^{ij}$ in Eq.~(\ref{eq:kij}) obtained with the Boltzmann simulation and the Eddington tensor $k^{ij}_{M1}$ estimated with the M1 approximation as in Eq.~(\ref{eq:kijM1}).
Note that both $k^{ij}$ and $k^{ij}_{M1}$ are measured in the laboratory frame throughout this paper.

Figure \ref{fig:Eddington} shows the typical radial distributions of diagonal (upper panels) and off-diagonal (lower panels) components of the Eddington tensors $k^{ij}$ and $k^{ij}_{M1}$ at $\epsilon_\mathrm{FR}=\langle\epsilon_\mathrm{FR}\rangle$ for $\nu_e$ at $t=10$ ms in 1D (left panels), 2D (middle panels), and 3D (right panels) simulations.
Here the average energy is defined as
\begin{eqnarray}
\langle\epsilon\rangle=\frac{\mathcal{E}}{\mathcal{N}},
\end{eqnarray}
where $\mathcal{N}$ is the number density in Eq.(\ref{eq:N}),
and $\mathcal{E}$ is the energy density given as
\begin{eqnarray}
\mathcal{E} = \int \epsilon^2 d\epsilon \int d\Omega_\nu \  \epsilon f (\epsilon, \Omega_\nu). \label{eq:E}
\end{eqnarray}
The integration above is done in the fluid-rest frame to calculate $\epsilon_\mathrm{FR}$.
The Eddington tensors obtained in the Boltzmann simulations and in the M1 closure approximation are drawn with the solid and dashed lines, respectively.
At $t=10$ ms, the radial distributions of diagonal components of $k^{ij}$ at $\epsilon_\mathrm{FR}=\langle\epsilon_\mathrm{FR}\rangle$ are almost the same among the 1D, 2D, and 3D cases (Fig.~\ref{fig:Eddington} (a), (b), and (c), respectively).
They are almost $1/3$ in the optically thick region extending up to 60 km.
They start to deviate from $1/3$ around the shock wave with $k^{rr}$ decreasing initially around 60 km and then increasing monotonically with radius.
The other diagonal components have the opposite trend.
On the other hand, the off-diagonal components of $k^{ij}$ are quite different among the 1D, 2D and 3D cases (Fig.~\ref{fig:Eddington}(d), (e), and (f), respectively).
All components in 1D and $k^{r\phi}$ and $k^{\theta\phi}$ in 2D are zero identically due to the symmetry of the neutrino distribution function in momentum space (Fig.~\ref{fig:NeutrinoDistribution} (d) and (e)).
Only $k^{r\theta}$ in 2D and all off-diagonal components in 3D have nonvanishing values.

Now we compare the Boltzmann Eddington tensor $k^{ij}$ with the M1 counterpart $k^{ij}_{M1}$.
The diagonal components $k^{ii}$ differ from $k^{ii}_{M1}$ around the shock wave at $t=10$ ms (Fig.~\ref{fig:Eddington} (a), (b), and (c)).
In fact, the latters change montonically with radius.
It is also found that the absolute value of $k^{\theta\phi}_{M1}$ tends to be smaller than those of other off-diagonal components $k^{r\theta}_{M1}$ and $k^{r\phi}_{M1}$ both in the transition and optically thin regions (Fig.~\ref{fig:Eddington} (f)).
This is understandable if one recalls that the radial flux dominates over other components in these regions.
We find a large difference between $k^{\theta\phi}$ and $k^{\theta\phi}_{M1}$ in the same regions (Fig.~\ref{fig:Eddington} (f)).
This suggests that the interpolation between the optically thick and thin limits employed in the M1 prescription is not very good for this component in the 3D simulation.

\section{PRINCIPAL-AXES ANALYSIS OF THE EDDINGTON TENSOR \label{sec:analysis}}

\begin{figure*}[ht!]
\begin{center}
\includegraphics[width=\hsize]{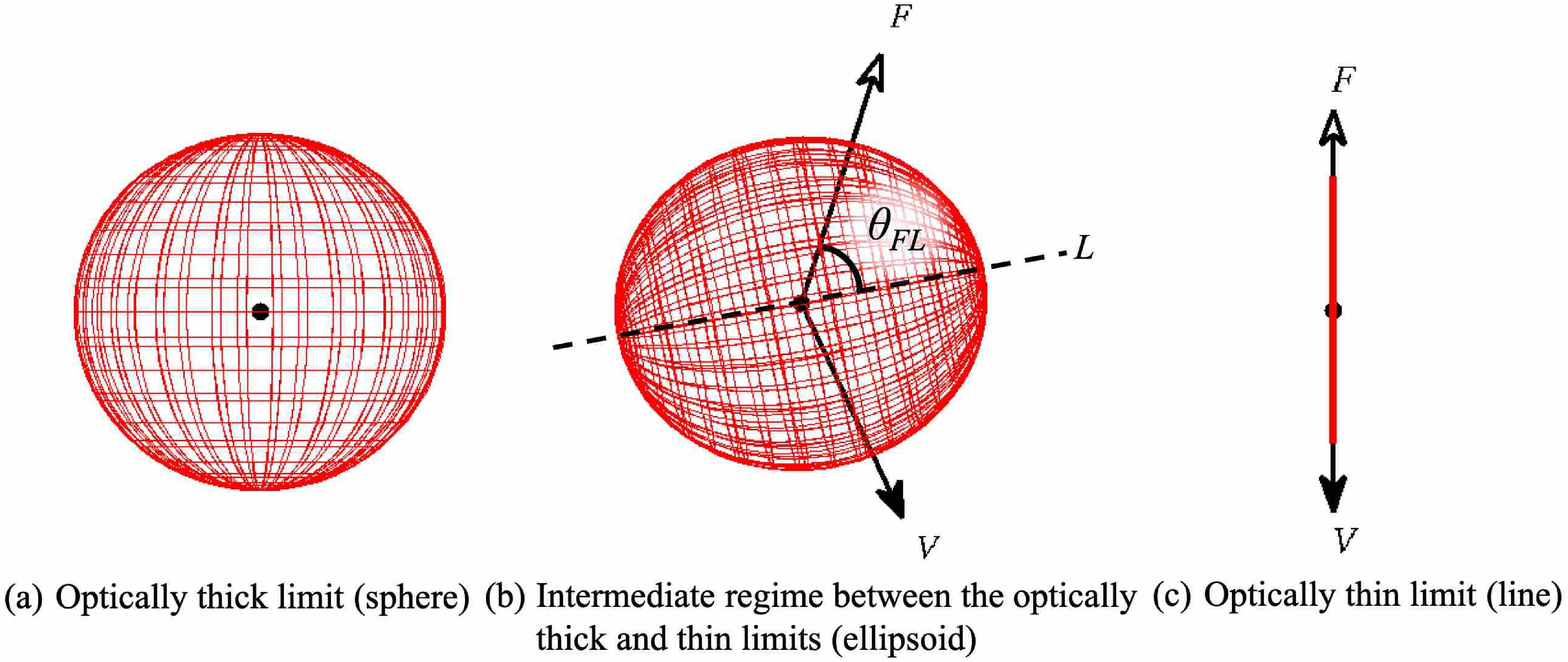}
\end{center}
\caption{Three representative configurations of the ellipsoid in the optically thick limit (left panel), the intermediate regime between the optically thick and thin limits (middle panel), and the optically thin limit (right panel).
The longest principal axis $L$ is denoted by a dashed line.
The directions of the neutrino energy flux $F$ and the matter velocity $V$ are represented by the arrows with white and black heads, respectively.
The angle between $F$ and $L$ is designated as $\theta_{FL}$.
\label{fig:ellipsoid}}
\end{figure*}

In this section, we apply a new analysis in the comparison of the Eddington tensors obtained from the Boltzmann simulations and the M1 closure approximation.
The Eddington tensor $k^{ij}$ in Eq.~(\ref{eq:kij}) can be written as
\begin{align}
    \mathbf{k}
    &=k^{rr}(\mathbf{e}_r\otimes\mathbf{e}_r)
    +k^{r\theta}(\mathbf{e}_r\otimes\mathbf{e}_\theta)
    +k^{r\phi}(\mathbf{e}_r\otimes\mathbf{e}_\phi) \nonumber\\
    &+k^{\theta r}(\mathbf{e}_\theta\otimes\mathbf{e}_r)
    +k^{\theta\theta}(\mathbf{e}_\theta\otimes\mathbf{e}_\theta)
    +k^{\theta\phi}(\mathbf{e}_\theta\otimes\mathbf{e}_\phi) \nonumber\\
    &+k^{\phi r}(\mathbf{e}_\phi\otimes\mathbf{e}_r)
    +k^{\phi\theta}(\mathbf{e}_\phi\otimes\mathbf{e}_\theta)
    +k^{\phi\phi}(\mathbf{e}_\phi\otimes\mathbf{e}_\phi),
\end{align}
where $\mathbf{e}_r$, $\mathbf{e}_\theta$, and $\mathbf{e}_\phi$ denote the basis vectors of the spherical coordinate system.
Since $k^{ij}$ is a real symmetric tensor, it has real eigenvalues and eigenvectors, which are easily obtained with the Jacobi method described in Appendix \ref{sec:jacobi}.
The diagonalized tensor $\mathbf{D}$ of $\mathbf{k}$ in Eq.~(\ref{eq:D}) is expressed as
\begin{align}
\mathbf{D}
    &=\lambda^{1}(\mathbf{e}_1\otimes\mathbf{e}_1)
    +\lambda^{2}(\mathbf{e}_2\otimes\mathbf{e}_2)
    +\lambda^{3}(\mathbf{e}_3\otimes\mathbf{e}_3),
\end{align}
where $\lambda^j$ and $\mathbf{e}_j$ denote the $j$-th eigenvalue and the corresponding normalized eigenvector of $\mathbf{k}$, respectively.  
Then the rotation matrix $\mathbf{V}$ in Eq.~(\ref{eq:V}) is defined as
\begin{align}
\mathbf{V}
    &=V^{r1}(\mathbf{e}_r\otimes\mathbf{e}_1)
    +V^{r2}(\mathbf{e}_r\otimes\mathbf{e}_2)
    +V^{r3}(\mathbf{e}_r\otimes\mathbf{e}_3) \nonumber\\
    &+V^{\theta 1}(\mathbf{e}_\theta\otimes\mathbf{e}_1)
    +V^{\theta 2}(\mathbf{e}_\theta\otimes\mathbf{e}_2)
    +V^{\theta 3}(\mathbf{e}_\theta\otimes\mathbf{e}_3) \nonumber\\
    &+V^{\phi 1}(\mathbf{e}_\phi\otimes\mathbf{e}_1)
    +V^{\phi 2}(\mathbf{e}_\phi\otimes\mathbf{e}_2)
    +V^{\phi 3}(\mathbf{e}_\phi\otimes\mathbf{e}_3),
\end{align}
where $(V^{rj}, V^{\theta j}, V^{\phi j})$ are the $r$-, $\theta$- and $\phi$-components of the $j$-th eigenvector.
Using these eigenvalues and eigenvectors, we visualize the Eddington tensor as an ellipsoid in momentum space,
\begin{equation}
\left(\frac{\mathbf{p}\cdot \mathbf{e}_1}{\lambda^1}\right)^2 + \left(\frac{\mathbf{p} \cdot \mathbf{e}_2}{\lambda^2}\right)^2 + \left(\frac{\mathbf{p} \cdot \mathbf{e}_3}{\lambda^3}\right)^2 = 1,
\end{equation}
where $\mathbf{p}$ is the momentum vector. 
Note that the $\mathrm{P_X}$-, $\mathrm{P_Y}$-, and $\mathrm{P_Z}$-axes in momentum space are parallel to the basis vectors $\mathbf{e}_\theta$,  $\mathbf{e}_\phi$, and $\mathbf{e}_r$, respectively (Fig.~\ref{fig:coordinate}).
The surface of the triaxial ellipsoid is also represented parametrically as
\begin{equation}
\left(\begin{array}{c}
      p^x\\
      p^y\\
      p^z
      \end{array}
\right)=
\left(\begin{array}{ccc}
      V^{\theta 1} & V^{\theta 2} & V^{\theta 3} \\
      V^{\phi 1} & V^{\phi 2} & V^{\phi 3}\\ 
      V^{r 1} & V^{r 2} & V^{r 3}
\end{array}
\right)
\left(\begin{array}{l}
      \lambda^1 \cos u \cos v \\
      \lambda^2 \cos u \sin v \\
      \lambda^3 \sin u 
      \end{array}
\right), \label{eq:ellipsoid}
\end{equation}
where $(p^x, p^y, p^z)$ are the coordinates of a point on the ellipsoidal surface and $u$ and $v$ are the parameters.
Of the three eigenvalues, $\lambda^3$ is chosen to be the largest, representing the longest axis of the ellipsoid.

\begin{figure*}[ht!]
\begin{center}
\includegraphics[width=18cm]{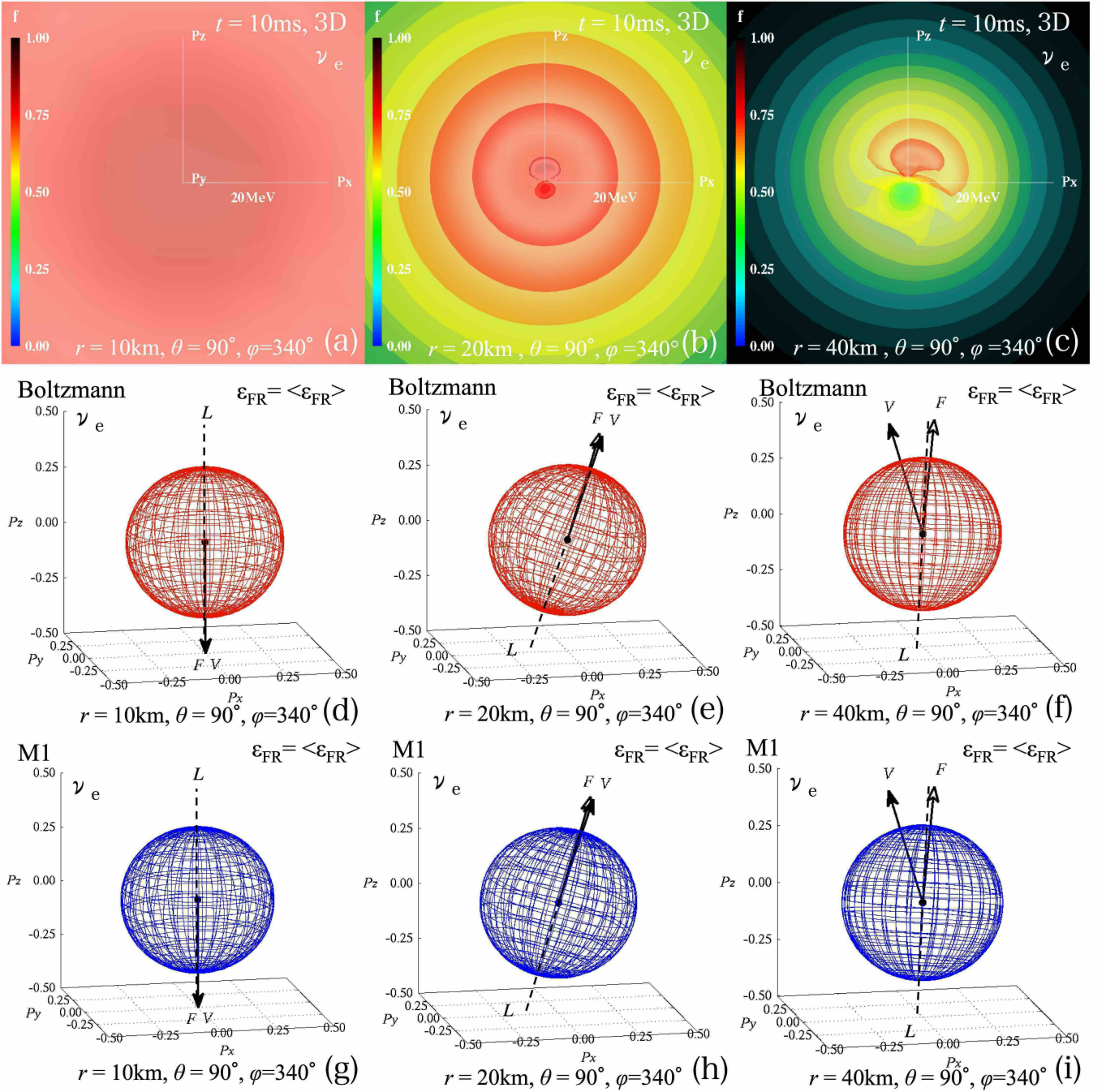}
\end{center}
\caption{The distribution functions $f$ of $\nu_e$ in momentum space (upper panels), and the corresponding ellipsoid of the Eddington tensors obtained in the Boltzmann simulation (middle panels) and those evaluated in the M1 closure approximation (lower panels)  for $\epsilon_\mathrm{FR}=\langle\epsilon_\mathrm{FR}\rangle$ at $r=$10 km (left panels), 20 km (middle panels), and 40 km (right panels).
\label{fig:NeutrinoEllipsoid}}
\end{figure*}

\begin{figure*}[ht!]
\begin{center}
\includegraphics[width=18cm]{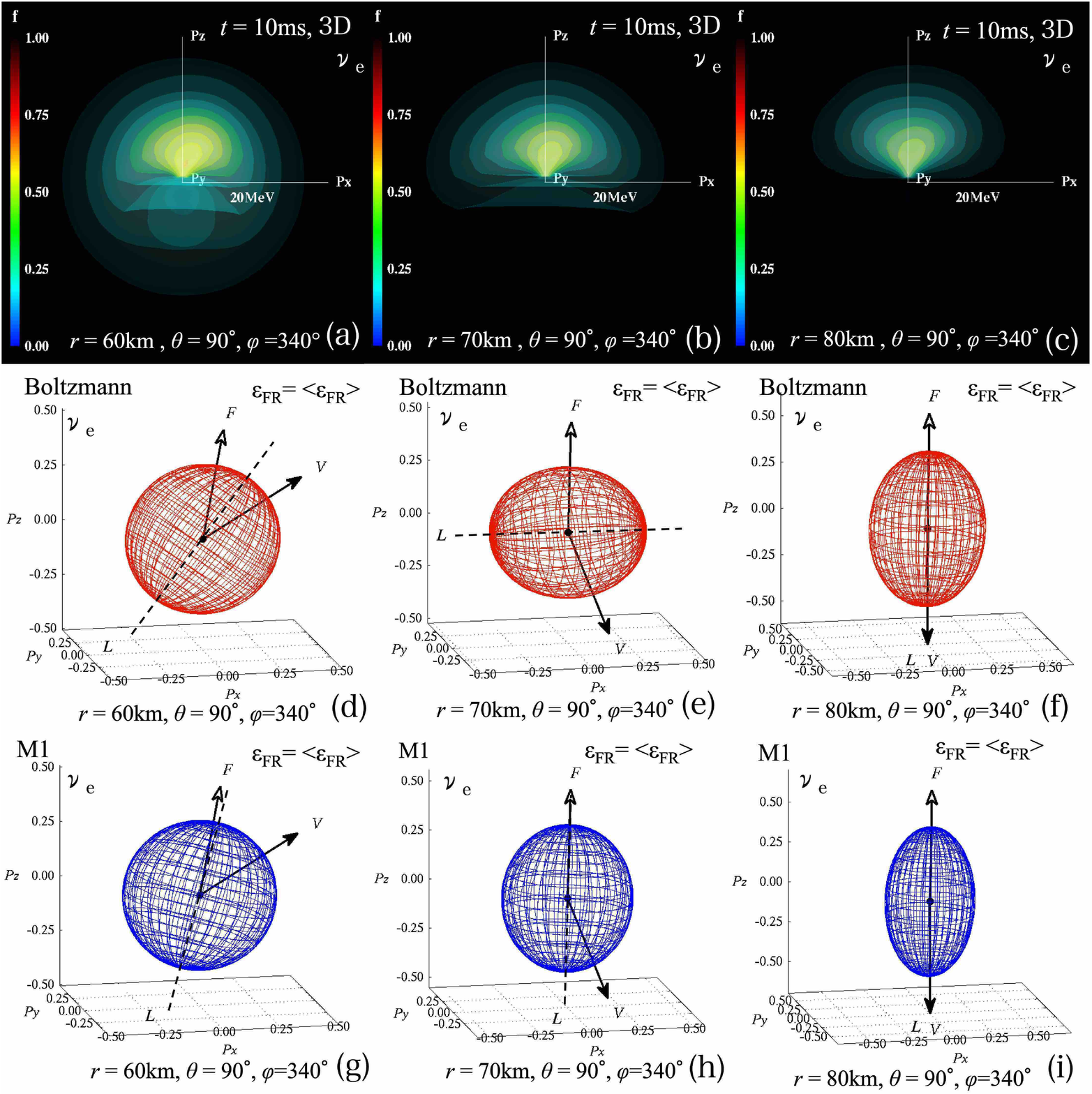}
\end{center}
\caption{The same as the previous figure except for the radius at 60 km (left panels), 70 km (middle panels), and 80 km (right panels).
\label{fig:NeutrinoEllipsoid2}}
\end{figure*}

Figure~\ref{fig:ellipsoid} shows three representative configurations of the ellipsoid.
In the optically thick limit, the ellipsoid is reduced to a sphere for the vanishing matter velocity (Fig.~\ref{fig:ellipsoid} (a)).
Recall that we are working in the laboratory frame.
The Eddington tensor has a single three-fold degenerate eigenvalue and three mutually orthogonal eigenvectors, the orientation of which is arbitrary.
In this case, the Jacobi method returns the eigenvectors as
\begin{equation}
\left(\begin{array}{ccc}
      V^{\theta 1} & V^{\theta 2} & V^{\theta 3} \\
      V^{\phi 1} & V^{\phi 2} & V^{\phi 3}\\ 
      V^{r 1} & V^{r 2} & V^{r 3}
\end{array}
\right)=
\left(\begin{array}{ccc}
      1 & 0 & 0 \\
      0 & 1 & 0\\ 
      0 & 0 & 1
\end{array}
\right).
\label{eq:Jacobi_thick}
\end{equation}
Note, however, that the completely isotropic distribution is never realized in the supernova core owing to matter inhomogeneity, which would induce diffusion of neutrinos even in the optically thick region.
In the M1 closure approximation applied in the fluid-rest frame, the situation is the same, with the dominant isotropic term of $\gamma^{ij}/3$ in Eq.~(\ref{eq:Pthick}) being modified by the minor term of $F^i F^j/|F|^2$ in Eq.~(\ref{eq:Pthin}). 
It is noted that in this approximation $F$ gives the eigenvector corresponding to the largest eigenvalue in the almost all region for $r\gtrsim 10$km at $t=10$ ms post bounce in this study (see Appendix~\ref{sec:analysisA}).
In the optically thin limit, on the other hand, the ellipsoid is reduced to a line  (Fig.~\ref{fig:ellipsoid} (c)).
There is only one non-vanishing eigenvalue.
The corresponding eigenvector gives the direction of the neutrino energy flux also in this case.
It is obvious that there is a two-fold degenerate vanishing eigenvalue in addition to the unique non-vanishing eigenvalue.
Finally, in the intermediate regime between the optically thick and thin limits, the ellipsoid is triaxial in general (Fig.~\ref{fig:ellipsoid} (b)).
The longest principal axis is denoted by $L$ hereafter.
The directions of the neutrino energy flux $F$ and the matter velocity $V$ are also represented by the arrows with white and black heads, respectively, in the figure.
The angle between $F$ and $L$ is designated as $\theta_{FL}$, and its cosine $\mu_{FL} = \cos\theta_{FL}$ is used for the later analysis.

\begin{figure*}[ht!]
\begin{center}
\includegraphics[width=\hsize]{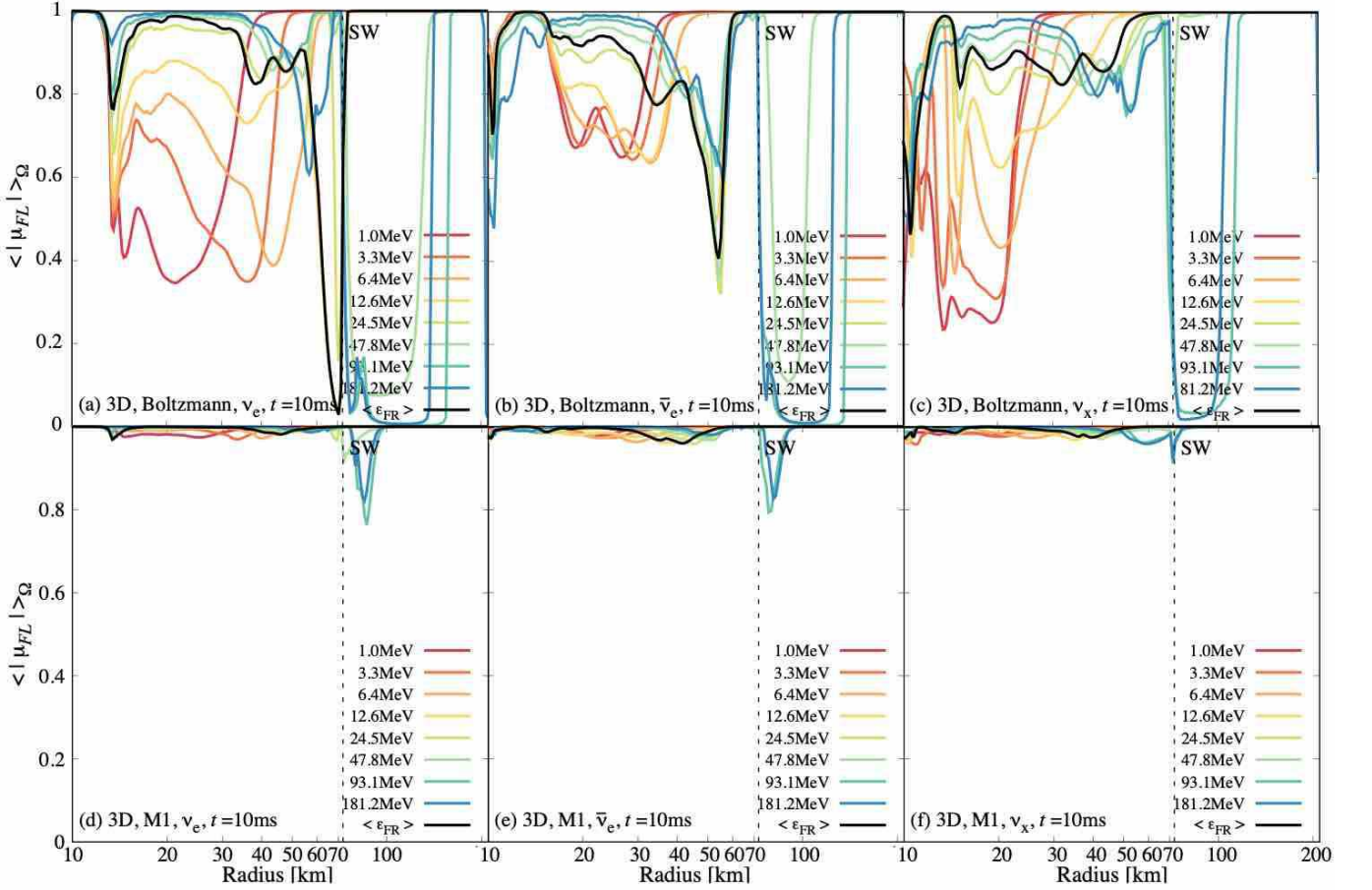}
\end{center}
\caption{The radial profile of $\langle|\mu_{FL}|\rangle_\Omega$ for $\nu_e$ (left panels), $\bar{\nu}_e$ (middle panels), and $\nu_x$ (right panels), where $\langle|\mu_{FL}|\rangle_\Omega$ denotes the absolute value of $\mu_{FL}$ averaged over the whole solid angle $\Omega$ at each radius.
It is unity (zero) if $L$ is parallel (perpendicular) to $F$.
The vertical dashed line indicates the radial position of the shock wave.
\label{fig:Inner1D}}
\end{figure*}

Figure \ref{fig:NeutrinoEllipsoid} shows the distribution functions $f$ of $\nu_e$ in momentum space (upper panels) and the corresponding ellipsoids of the Eddington tensors obtained in the Boltzmann simulation (middle panels) and those evaluated in the M1 approximation (lower panels) at three radii from 10 km to 40 km for the neutrino energy $\epsilon_\mathrm{FR}=\langle\epsilon_\mathrm{FR}\rangle$, i.e., the average energy.
At $r=10$ km, the neutrino distribution in momentum space is almost isotropic at all energies in the fluid rest frame owing to the tight coupling with matter and, as a result, $F$ is nearly parallel to $V$ in the laboratory frame (Fig.~\ref{fig:NeutrinoEllipsoid} (d)).
The Eddington tensor that is evaluated in the laboratory frame is also slightly elongated in the direction of $F$.
With the increasing radius, high-energy neutrinos are depleted (Fig.~\ref{fig:NeutrinoEllipsoid} (b)) and, more importantly, anisotropy becomes pronounced particularly for low-energy neutrinos
(Fig.~\ref{fig:NeutrinoEllipsoid} (c));
it is observed at the same time (Fig. \ref{fig:NeutrinoEllipsoid} (e) and (f)) that $F$ and $V$ start to be misaligned with each other in the convective region ($r\sim40$ km) as the interactions of neutrinos with matter get weaker and neutrinos diffuse out according to the local gradient of the neutrino number density.
It is noted that the ellipsoids for the Boltzmann simulation  (Fig.~\ref{fig:NeutrinoEllipsoid} (d)-(f)) agree well with those for the M1 approximation (Fig.~\ref{fig:NeutrinoEllipsoid} (g)-(i)), with the longest axis of the ellipsoid $L$ being almost aligned with the direction of the energy flux $F$.

\begin{figure*}[ht!]
\begin{center}
\includegraphics[width=\hsize]{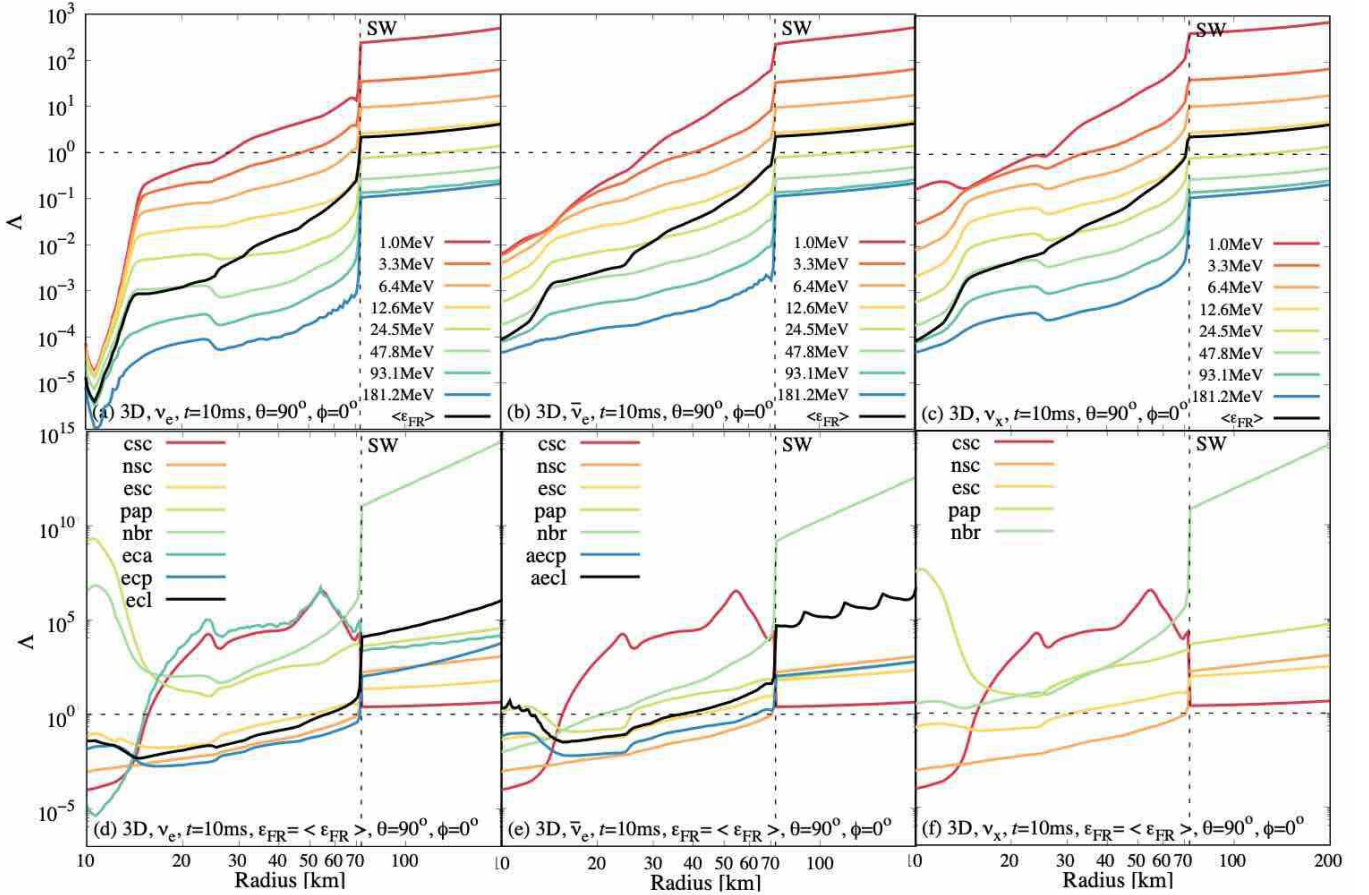}
\end{center}
\caption{The radial profile of $\Lambda$ for $\nu_e$ (left panels), $\bar{\nu}_e$ (middle panels), and $\nu_x$ (right panels), where $\Lambda$ is the mean free path normalized by the radius.
The shock position is indicated by the vertical thick dashed line.
The horizontal thin dashed line corresponds to $\Lambda = 1$.
\label{fig:MFP}}
\end{figure*}

Figure \ref{fig:NeutrinoEllipsoid2} shows the same quantities as Figure \ref{fig:NeutrinoEllipsoid} does except for the spatial positions considered: 60 km to 80 km.
At these larger distances from the center of the core, the neutrinos are more concentrated toward the $+\mathrm{P_Z}$ axis in momentum space (Fig.~\ref{fig:NeutrinoEllipsoid2} (a)-(c)).
The energy flux $F$ and the matter velocity $V$ tend to the $+\mathrm{P_Z}$ and $-\mathrm{P_Z}$ axes, respectively, with the increasing radius (Fig.~\ref{fig:NeutrinoEllipsoid2} (d)-(f)).
The ellipsoid changes it’s shape from a near sphere (Fig.~\ref{fig:NeutrinoEllipsoid2} (d)) to a horizontally elongated ellipsoid (Fig.~\ref{fig:NeutrinoEllipsoid2} (e)), and finally to a vertically extended ellipsoid (Fig.~\ref{fig:NeutrinoEllipsoid2} (f)),
where the terms “horizontal" and “vertical" mean “perpendicular" and “parallel" to the direction of $F$, respectively.
The horizontally elongated configuration observed above is consistent with the fact that $k^{rr}$ $(k^{\theta \theta}, k^{\phi \phi})$ 
is lower (higher) than 1/3 there (Fig.~\ref{fig:Eddington} (a)-(c)), which can be also seen in the results of the neutrino radiation transport simulations solving the Boltzmann equation with $S_N$ method \citep{Smit2000} and with Monte-Carlo approach \citep{Janka1992, Murchikova2017}.
It occurs not only in
3D but also in 1D and 2D simulations.
In the bottom panels of Fig.~\ref{fig:NeutrinoEllipsoid2},
we exhibit the ellipsoids given by the M1 closure approximation.
At 60 km, the directions of the longest principal axis $L$ are different between the Boltzmann simulation and the M1 approximation although the ellipsoids are almost spherical in both cases (Fig.~\ref{fig:NeutrinoEllipsoid2} (d) and (g)).
At $r = 70$ km, however, the difference is more substantial.
In fact, the ellipsoid in the M1 approximation is extended vertically whereas it is wider horizontally in the Boltzmann simulation (Fig.~\ref{fig:NeutrinoEllipsoid2} (e) and (h)). 
The difference remains even at $r = 80$ km, where the M1 approximation gives a more elongated ellipsoid although $L$ in the Boltzmann simulation is now aligned with $F$ (Fig.~\ref{fig:NeutrinoEllipsoid2} (f) and (i)): $k^{rr}_{M1} > k^{rr}$, $k^{\theta\theta}_{M1} < k^{\theta\theta}$, and $k^{\phi\phi}_{M1} < k^{\phi\phi}$ at $r>70$ km (Fig.~\ref{fig:Eddington} (a)-(c)), which is also common to the 1D, 2D, and 3D simulations.
Note that to what extent these are the case is rather sensitive to the numerical resolution (see Appendix~\ref{sec:resolution}).
We stress again, however, that $L$ is always almost aligned with $F$ in the M1 approximation (Fig.~\ref{fig:NeutrinoEllipsoid2} (g), (h), and (i)), while it is not necessarily the case in reality (Fig.~\ref{fig:NeutrinoEllipsoid2} (d) and (e)).

Figure \ref{fig:Inner1D} shows the radial profile of  $\langle|\mu_{FL}|\rangle_\Omega$ for $\nu_e$ (left panels), $\bar{\nu}_e$ (middle panels), and $\nu_x$ (right panels), where $\langle|\mu_{FL}|\rangle_\Omega$ denotes the absolute value of $\mu_{FL}$ averaged over the whole solid angle $\Omega$ at each radius.
It is unity (zero) if $L$ is parallel (perpendicular) to $F$.
The vertical dashed line in the figure indicates the radial position of the shock wave.
The prompt convection prevails from $r\sim10$ km to 60 km while a laminar shocked flow exists from 60 km to 70 km, and there is a supersonic accretion from 70 km to 200 km.
The results for $\nu_e$ in the Boltzmann simulation show that in the convective region, $\langle|\mu_{FL}|\rangle_\Omega$ gets small $\sim 0.5$ for low-energy neutrinos (reddish lines) while for the average-energy neutrinos (yellow and green lines) it is deviated from unity substantially in the laminar region; in the supersonic accretion flow high-energy neutrinos (bluish lines) are affected.
The results for $\bar{\nu}_e$ and $\nu_x$ are more or less similar to those for $\nu_e$ although the dip in the laminar region is less remarkable for the average-energy neutrinos (Fig.~\ref{fig:Inner1D} (b) and (c)).
The results for the M1 closure approximation are quite different, on the other hand.
The longest principal axis of the ellipsoid obtained from $k^{ij}_{M1}$ is roughly parallel to the neutrino energy flux over the entire region, except just above the shock wave (Fig.~\ref{fig:Inner1D} (d)-(f)).
This is due to the fact that the term of $F^iF^j/F^2$ in Eq.~(\ref{eq:Pthin}) have a dominant role to determine the orientation of the ellipsoid in the M1 closure approximation even in the laboratory frame.
As a result, the M1 prescription Eq.~(\ref{eq:zeta}) cannot reproduce the horizontally wide ellipsoids, which are realized in different regions depending on the neutrino energy.

\begin{figure*}[ht!]
\begin{center}
\includegraphics[width=\hsize]{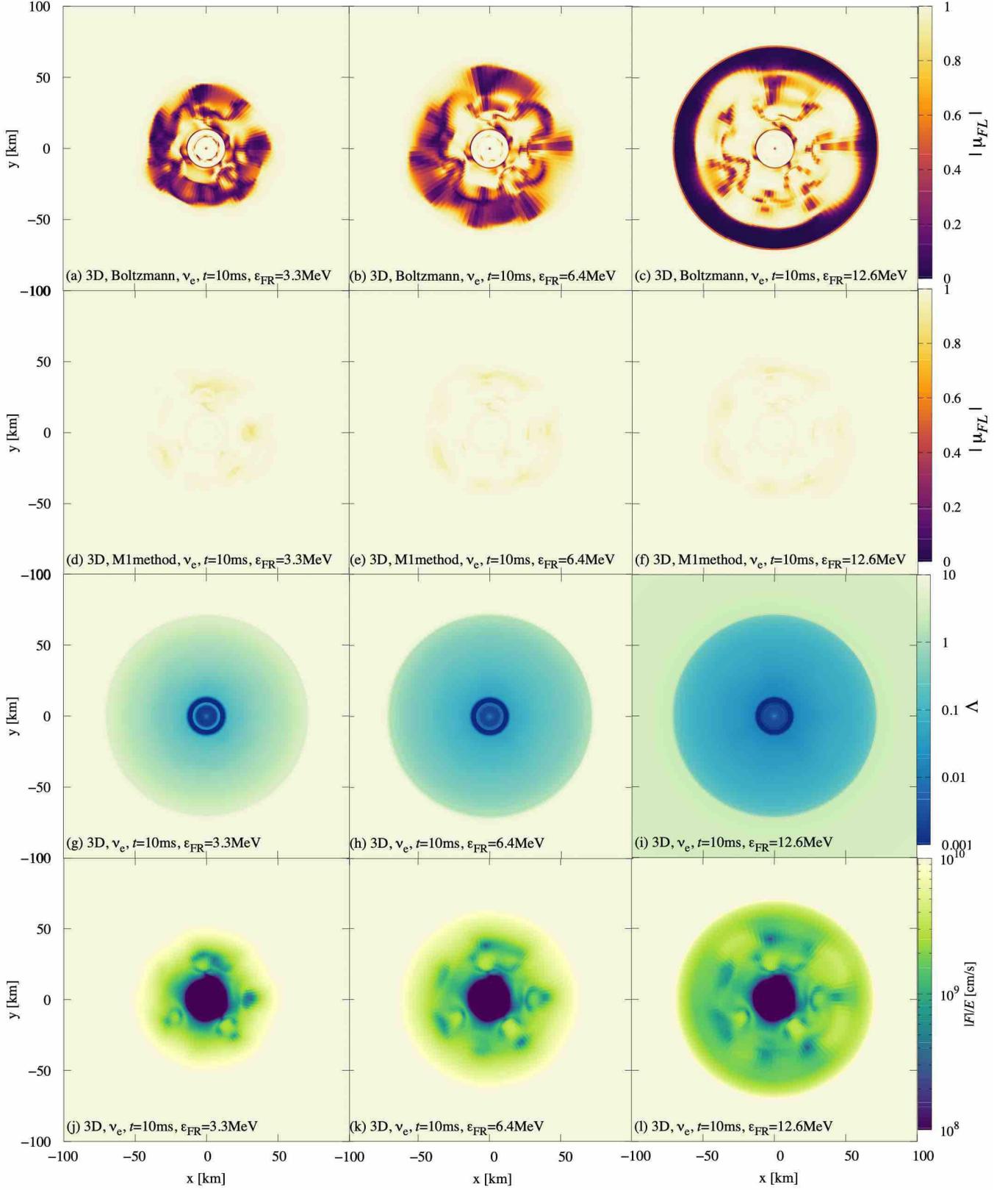}
\end{center}
\caption{The color maps of $|\mu_{FL}|$ for the Boltzmann simulation (1st rows) and the M1 closure approximation (2nd rows), $\Lambda$ (3rd rows), and $|F|/E$ (4th rows) on the equatorial plane for $\nu_e$ at $t=10$ ms.
The left, middle, and right panels are the results for $\epsilon_\mathrm{FR}$ = 3.3, 6.4, and 12.6 MeV, respectively.
\label{fig:Inner_MFP_Flux_merid}}
\end{figure*}

\begin{figure*}[ht!]
\begin{center}
\includegraphics[width=\hsize]{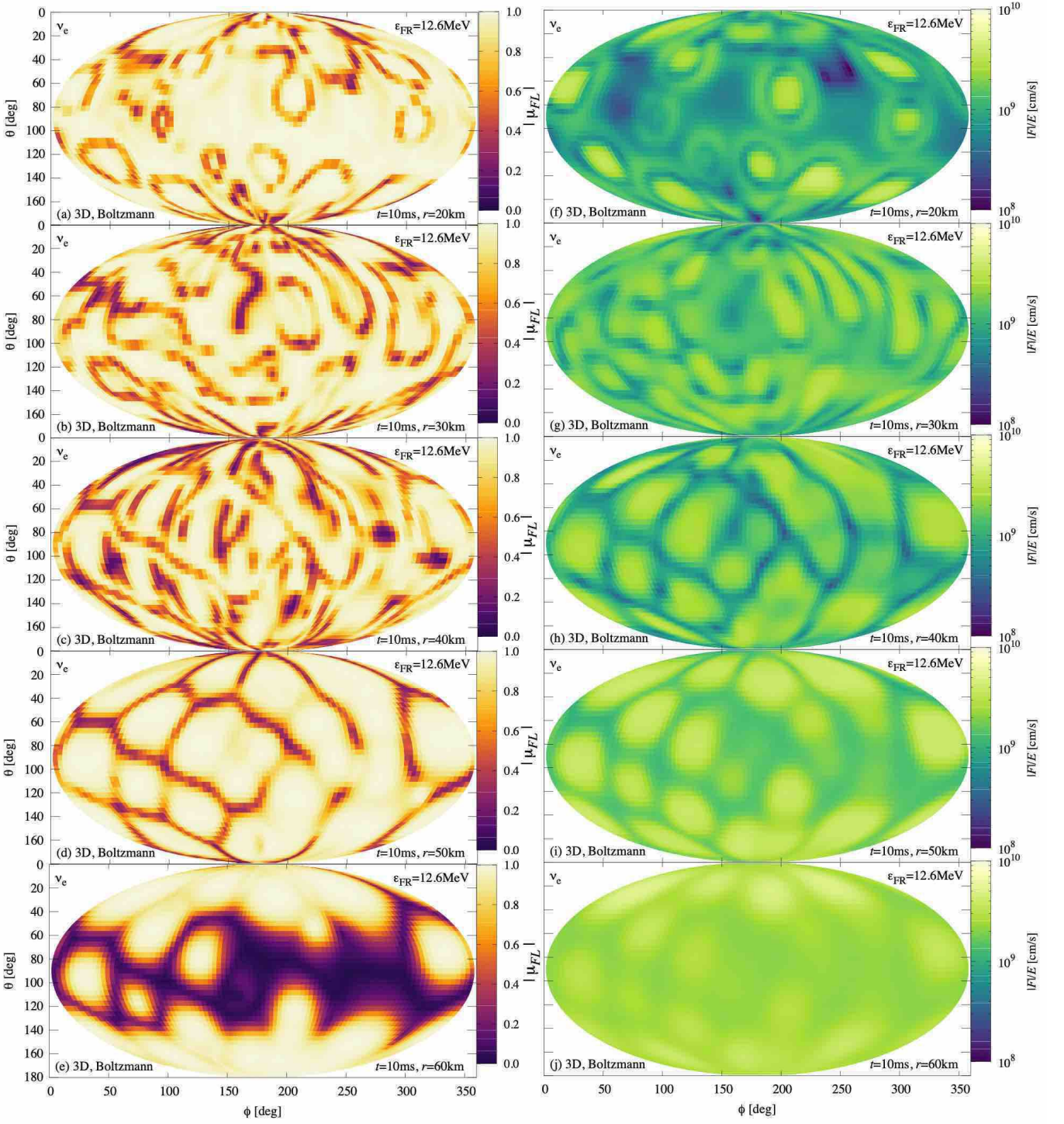}
\end{center}
\caption{The Mollweide projection map of $|\mu_{FL}|$ (left panels) and $|F|/E$ (right panels) for $\nu_e$ at various radii for $\epsilon_\mathrm{FR}$=12.6 MeV at $t=10$ms.
The inhomogeneous distribution of $|\mu_{FL}|$ at $r=60$ km is responsible for the rapid growth of the prompt convection around the pole axis due to the fine mesh around the coordinate singularity.
\label{fig:Inner_Flux_mollwide_en007}}
\end{figure*}

\begin{figure*}[ht!]
\begin{center}
\includegraphics[width=\hsize]{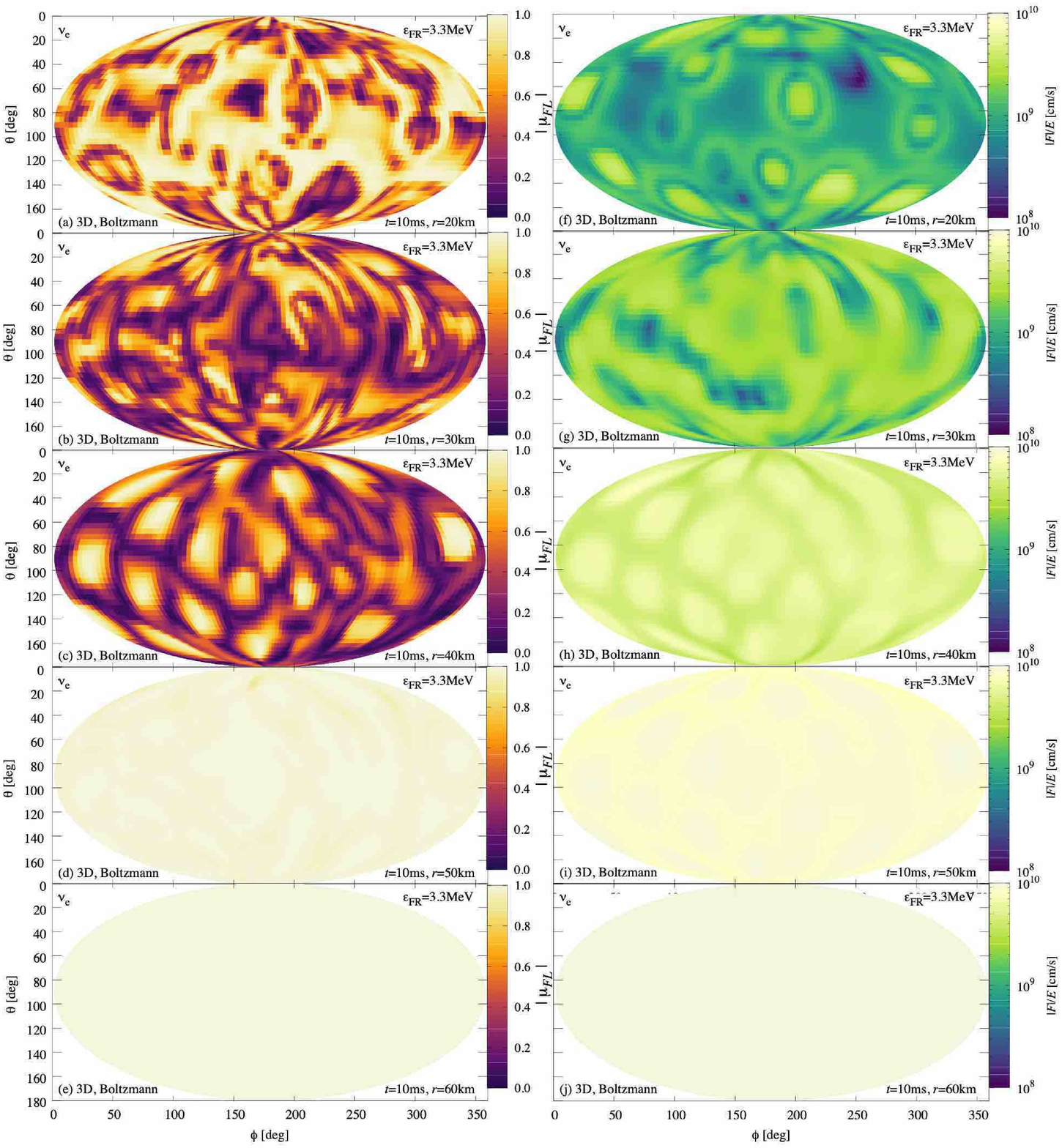}
\end{center}
\caption{The same as the previous figure except for the energy $\epsilon_\mathrm{FR}$=3.3 MeV.
\label{fig:Inner_Flux_mollwide_en003}}
\end{figure*}

Figure \ref{fig:MFP} shows the radial profile of $\Lambda$ for $\nu_e$ (left panels), $\bar{\nu}_e$ (middle panels), and $\nu_x$ (right panels), where $\Lambda$ is the mean free path  normalized by the radius.
The shock position is indicated by the vertical thick dashed line whereas the horizontal thin dashed line corresponds to $\Lambda = 1$.
In the bottom row of Fig.~\ref{fig:MFP}, 
we show the radial profiles of the contributions to $\Lambda$ of various interactions for the average neutrino energy (Fig.~\ref{fig:MFP} (d)-(f)).
In both of the convective and laminar shocked regions, the electron capture (ecp) gives the smallest $\Lambda$ for $\nu_e$ while the scattering with free nucleons (nsc) is dominant for $\bar{\nu}_e$ and $\nu_x$.
In the supersonic accretion flow, on the other hand, the coherent scattering with heavy nuclei (csc) determines $\Lambda$ for all neutrino species.
The region with $\Lambda \lesssim 1$ is the transition region and that is exactly where the approximation for neutrino transport should be validated by the Boltzmann simulation.
For low-energy neutrinos (reddish lines), $\Lambda \lesssim 1$ in the prompt convection region from 10 km to 60 km. For the neutrinos around the average energy (yellow and green lines) the region with $\Lambda \lesssim 1$ extends itself over both the prompt convection region and the laminar shocked region.
The value of $\Lambda$ for high-energy neutrinos (bluish lines) is below unity in the entire region up to 200 km, including the supersonic region.
It is hence clear that the misalignment of  $F$ and $L$ occurs in their own transition regions for different energies of neutrinos.

One finds from Fig.~\ref{fig:NeutrinoEllipsoid2} (b) for $\epsilon_\mathrm{FR} = \langle\epsilon_\mathrm{FR}\rangle \sim 12.6$ MeV just behind the shock wave that the neutrino angular distribution becomes hemispheric, i.e. almost isotropic only in one hemisphere, and this is associated with the horizontally elongated ellipsoids in these regions discussed earlier.
The hemispheric distribution is understood intuitively as follows: outgoing neutrinos with $\theta_\nu < \pi/2$ are mostly generated in the deeper and hence optically thicker region while those going inward with $\theta_\nu > \pi/2$ are emitted or back-scattered in the outer optically thinner region; as the reaction rates are certainly larger for the former, we obtain the angular distribution with the latter neutrinos highly depleted.
As the mean free path increases, the angular distribution of $f$ in the neutrino-rich hemisphere becomes forward-peaked while $f$ in the opposite hemisphere gets more diminished.
The hemispheric distribution emerges typically at $\Lambda \sim 0.1$ at $t = 10$ ms.
It is noted that the maximum entropy distributions for Fermi-Dirac radiation \citep[e.g.][]{Janka1992, Cernohorsky1994}, making the normalized radiation pressure smaller than 1/3 depending on the occupation density and flux factor, can give a similar distribution as the hemispheric one, where the boundary between neutrino-rich and poor is located at $\mu_\nu < 0$.
The relation of our results with the maximum entropy distributions for Fermi-Dirac radiation will be discussed in upcoming paper.
It should be stressed that in multi-dimensional cases, these hemispheres do not exactly correspond to the outgoing or ingoing directions.
At the early post-bounce period we consider here, this is particularly clear for low-energy neutrinos. 
In Fig.~\ref{fig:NeutrinoDistribution} we find the typical hemispheric distribution as the orange surface for $\epsilon \lesssim 10$ MeV at $r = 30$ km.
Its shape is axisymmetric with respect to the local radial direction in 1D (Fig.~\ref{fig:NeutrinoDistribution} (a)).
In 2D and 3D (Fig.~\ref{fig:NeutrinoDistribution} (b) and (c)), it is distorted by the prompt convection and, as mentioned, the two hemispheres are inclined from the local radial direction although it remains plane-symmetric with respect to the $\mathrm{P_Z}$-$\mathrm{P_X}$ plane in 2D.
It is repeated that these hemispheric distributions are the main origin of the horizontally elongated ellipsoid, for which $L$ and $F$ are orthogonal to each other, and the M1 approximation failed to reproduce the Eddington tensor (Fig.~\ref{fig:NeutrinoEllipsoid2} (e)).

In the multi-dimensional cases, the region, in which $L$ is perpendicular to $F$, appears in space in a complex way.
Figure \ref{fig:Inner_MFP_Flux_merid} shows the color maps of unaveraged $|\mu_{FL}|$ on the equatorial plane for the Boltzmann simulation (1st rows) and the M1 closure approximation (2nd rows) as well as $\Lambda$ (3rd rows) and $|F|/E$ (4th rows) for $\nu_e$ at $t=10$ ms.
The left, middle, and right panels are the results for $\epsilon_\mathrm{FR}=3.3, 6.4$, and 12.6 MeV, respectively.
In the upper two rows for $|\mu_{FL}|$, the thick colored region is the place, where $L$ and $F$ are orthogonal to each other and the ellipsoid becomes horizontally wide.
As was observed in Fig.~\ref{fig:Inner1D}, the radius of the thick colored region gets larger with increasing energy in the Boltzmann simulation (Fig.~\ref{fig:Inner_MFP_Flux_merid} (a)-(c)), while there is no such region for the M1 closure approximation (Fig.~\ref{fig:Inner_MFP_Flux_merid} (d)-(f)).
At low neutrino energies, $\epsilon_\mathrm{FR}=3.3$ and 6.4 MeV, these regions are distributed in a patchy fashion in the prompt convection zone (Fig.~\ref{fig:Inner_MFP_Flux_merid} (a) and (b)) and the angle-average of $|\mu_{FL}|$ becomes $\langle |\mu_{FL}|\rangle_\Omega \gtrsim 0.3$ there (Fig.~\ref{fig:Inner1D} (a)).
For the average neutrino energy $\epsilon_\mathrm{FR}\sim$12.6 MeV (Fig.~\ref{fig:Inner_MFP_Flux_merid} (c)), on the other hand, it is extended uniformly in the laminar shocked flow and, as a result, $\langle |\mu_{FL}|\rangle_\Omega$ is close to zero even just behind the shock wave (Fig.~\ref{fig:Inner1D} (a)).
However, there are also pockets of regions in the prompt convection zone, in which $L$ and $F$ are not aligned (Fig.~\ref{fig:Inner_MFP_Flux_merid} (c)).
They appear correlated not with the color map of $\Lambda$ (Fig.~\ref{fig:Inner_MFP_Flux_merid} (i)) but with $|F|/E$ (Fig.~\ref{fig:Inner_MFP_Flux_merid} (l)).

Figures \ref{fig:Inner_Flux_mollwide_en007} and \ref{fig:Inner_Flux_mollwide_en003} are the Mollweide projection maps of $|\mu_{FL}|$ (left panels) and $|F|/E$ (right panels) at various radii for $\epsilon_\mathrm{FR}$=12.6 and 3.3 MeV, respectively, at $t=10$ ms.
In Fig.~\ref{fig:Inner_Flux_mollwide_en007} for $\epsilon_\mathrm{FR}$=12.6 MeV, we can confirm the correlation between the $|\mu_{FL}|$ and $|F|/E$, which we found in Fig.~\ref{fig:Inner_MFP_Flux_merid} (c) and (l) above, at $r = 20-50$ km, where $\Lambda$ is $\gtrsim$ 0.01.
It is also observed that the regions of small $|\mu_{FL}|$ ellipsoids roughly agree with those of local minima of $|F|/E$.
It turns out that in these regions, where $\Lambda \gtrsim$ 0.01 for $\epsilon_\mathrm{FR}$=12.6 MeV, the ellipsoids are almost spherical and slightly elongated in the direction perpendicular to $F$ (see Fig.~\ref{fig:NeutrinoEllipsoid} (d)-(f)).
For $\epsilon_\mathrm{FR} = 3.3$ MeV, the pattern in $|\mu_{FL}|$ also looks very similar to that in $|F|/E$ at $r=40$ km  (Fig.~\ref{fig:Inner_Flux_mollwide_en003} (c), (h)), where $\Lambda$ is $\lesssim 1$ (Fig.~\ref{fig:MFP} (a)).
Although the neutrino distribution starts to become forward-peaked, it is still hemispheric and the ellipsoid is horizontally wide in the regions of $|\mu_{FL}|\sim 0$, which again roughly agree with the regions of local minima of $|F|/E$.
This is not true, however, at smaller radii (Fig.~\ref{fig:Inner_Flux_mollwide_en003} (a), (b), (f), (g)).
Note that there are still many regions, where $|\mu_{FL}|$ is much smaller than 1 and the ellipsoid is horizontally elongated substantially.
They are located at $\Lambda \sim 0.1$.
It seems, however, that the convection makes situations much more complicated in 3D.
In fact, the neutrino distribution function at $\epsilon_\mathrm{FR}$=3.3 MeV is severely distorted by complex matter motions in these regions (see the orange-colored isosurfaces in Fig.~\ref{fig:NeutrinoDistribution} (c) and (f)), which makes the neutrino flux less correlated with the shape of ellipsoid.
If convective motions at $\Lambda \sim 0.1$ are responsible indeed, it may not be easy to predict the place where the horizontally-elongated ellipsoids are produced and make an appropriate modification to the M1 prescription. That will be a future work.

\section{DISCUSSIONS AND CONCLUSIONS \label{sec:conclusion}}

We have done a radiation-hydrodynamical simulation of core-collapse supernova for a 11.2$M_\odot$ progenitor model in three-dimensional space with the full Boltzmann neutrino transport until 20 ms after bounce.
The time-dependent six-dimensional Boltzmann equations for three species of neutrinos and the three-dimensional hydrodynamic equations with the monopole Newtonian self-gravity have been solved with Furusawa and Togashi's equation of state.
What we have done in this paper are summarized as follows.
\begin{enumerate}
\item
We have investigated the neutrino distributions in the three-dimensional space at this early post-bounce phase.
Multiple round vortices are generated in the prompt convection that sets in at $\sim 10$ ms and produce in turn three-dimensional structures of density, entropy, and electron fraction in space.
We have observed that neutrinos move along with matter in the optically thick region and starts to decouple from matter in the transition region between the optically thick and thin limits, and finally propagate outward in the optically thin region.

\item
We have confirmed that the neutrino angular distributions in momentum space have no symmetry for 3D, while they have axisymmetry with respective to the radial direction and reflection-symmetry with respective to the $\mathrm{P_Z}$-$\mathrm{P_X}$ plane for 1D and 2D, respectively; all the off-diagonal components of Eddington tensor are nonvanishing for 3D; and there are some differences between the Eddington tensors obtained from the Boltzmann simulation and from the M1 closure approximation.

\item
We have applied a new analysis based on the principal axis transformation for the Eddington tensor to better understand these differences.
We visualize the Eddington tensor as the ellipsoid whose principal axes are parallel to its eigenvectors, having the lengths proportional to the corresponding eigenvalues.
The ellipsoid is reduced to a sphere in the optically thick limit and to a line parallel to the energy flux in the optically thin limit.
In between it is a triaxial ellipsoid in general.
We have found that the ellipsoid obtained directly from the Boltzmann simulation sometimes becomes horizontally wide, that is, elongated in the direction perpendicular to the energy flux, in the transition regime between the optically thick and thin limits.
This is in sharp contrast to the M1 closure approximation, in which the longest principal axis of the ellipsoid is always almost parallel to the direction of the energy flux.

\item
The horizontally wide Eddington tensor emerges with high probabilities in the transition region between the optically thick and thin limits, where the mean free path divided by the radius is $\Lambda \sim 0.1$.
As a matter of fact, in the convective region, the horizontally wide ellipsoid tends to occur at places, where $\Lambda$ is a bit higher or lower than 0.1 and, in addition, the neutrino velocity $|F|/E$ takes locally minimum values.
We have observed, however, that convective matter motions make the situation more complicated.
This may make it difficult to improve the M1 approximation in these regions.
\end{enumerate}

This paper is the very first step in our project of 3D Boltzmann radiation-hydrodynamics simulations of core-collapse supernovae.
In fact, we have focused only on the prompt convection phase at $t\sim10$ ms post bounce and paid attention to the neutrino distributions in momentum space.
In this very early phase, the entire post-shock region is optically thick with $\Lambda < 1$ for neutrinos with $\gtrsim 10$MeV.
The convection region is hence located deep in the optically thick region for the average neutrinos.
At much later times, the neutrino-driven convection or the standing accretion shock instability (SASI) and even the lepton-driven convection occur in more optically thin regions.
We hence should examine how the Eddington tensor is affected by these multi-dimensional matter motions: in particular we are interested in whether the horizontally elongated ellipsoid occurs in a similar fashion and, more importantly, if it has an impact on the shock revival.
If it does, it will be important to try seriously to improve the M1 closure approximation.
Not to mention, our simulation is hardly perfect.
The numerical resolutions in space and momentum space, the monopolar Newtonian gravity and the one-dimensional treatment of the innermost region at $r<8$ km are certainly concerns, to mention a few.
We are planning to perform longer simulations with better resolutions on Fugaku computer, the next generation flagship supercomputer of Japan whose capability is expected to be more than 10 times as large as K-computer's, and will address these issues there.

\acknowledgments

This research used high-performance computing resources of the K-computer, FX100 provided by the Nagoya University ICTS, and Oakforest-PACS provided by JCAHPC through the HPCI System Research Project (Project ID: hp140211, 150225, 160071, 160211, 170031, 170230, 170304, 180111, 180179, 180239, 190100, 190160, 200102, 200124),
NEC SX Aurora Tsubasa at KEK,
Research Center for Nuclear Physics (RCNP) at Osaka University,
and the XC50 and the general common use computer system provided by CfCA in the National Astronomical Observatory of Japan (NAOJ).
Large-scale storage of numerical data is supported by JLDG constructed over SINET4 of NII. 
This work was supported in part by Grants-in-Aid for Scientific Research (26104006, 15K05093, 19K03837), and Grant-in-Aid for Scientific Research on Innovative areas ”Gravitational wave physics and astronomy: Genesis” (17H06357, 17H06365) and ”Unraveling the History of the Universe and Matter Evolution with Underground Physics” (19H05802) from the Ministry of Education, Culture, Sports, Science and Technology (MEXT), Japan. This work was also partly supported by research programs at K-computer of the RIKEN R-CCS, HPCI Strategic Program of Japanese MEXT, Priority Issue on Post-K-computer (Elucidation of the Fundamental Laws and Evolution of the Universe), Joint Institute for Computational Fundamental Sciences (JICFus), and Japan Society for the Promotion of Science (JSPS) Grant-in-Aid for Young Scientists(Start-up, JP19K23435).
A. H. was supported in part by MEXT Grant-in-Aid for Research Activity Start-up (19K23435).
S. F. was supported by JSPS KAKENHI (19K14723).

\appendix

\section{Resolution test \label{sec:resolution}}

\begin{figure}[ht!]
\begin{center}
\includegraphics[width=\hsize]{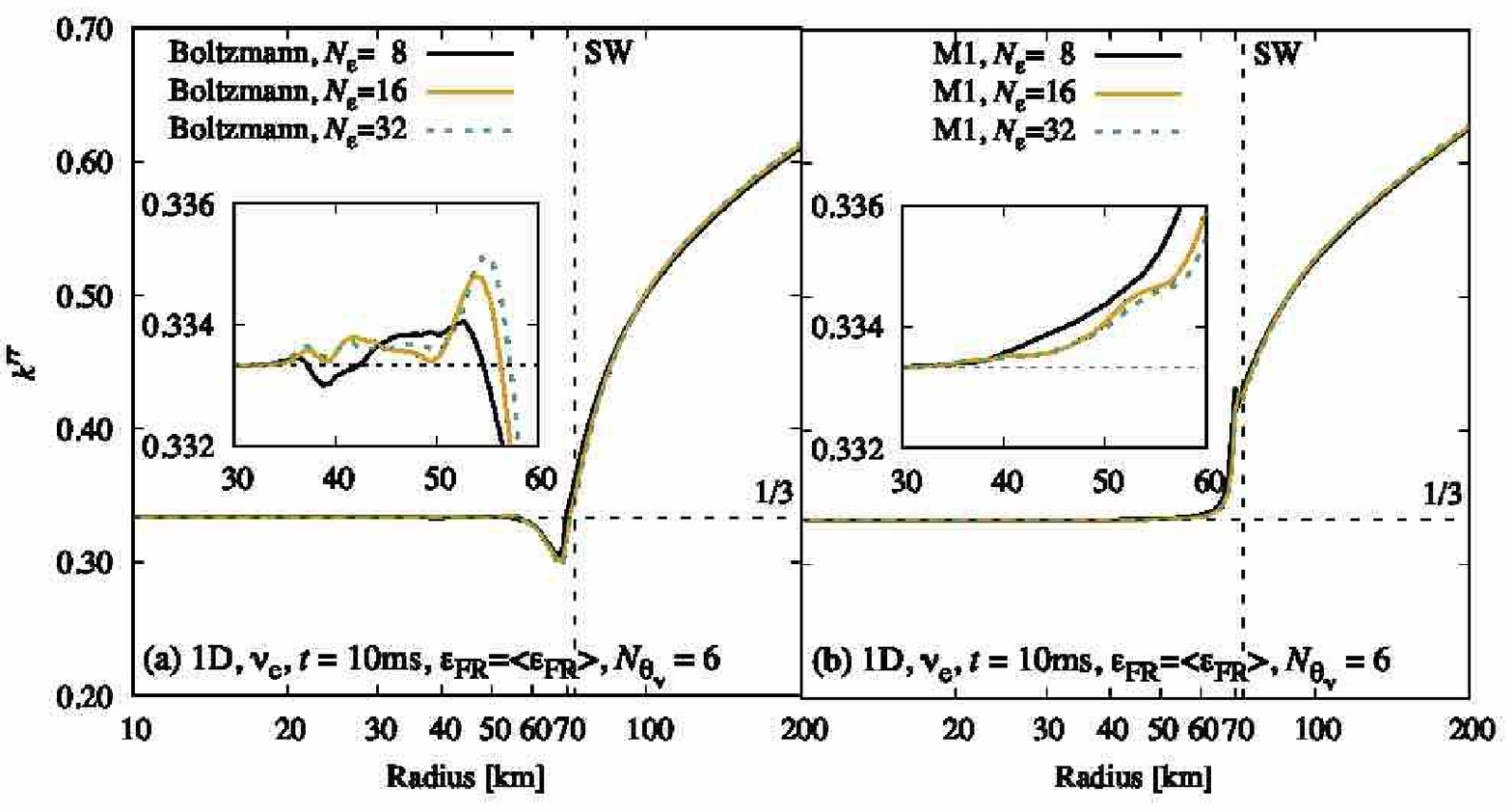}
\end{center}
\caption{The energy-resolution dependence of $k^{rr}$ at $\epsilon_\mathrm{FR}=\langle \epsilon_\mathrm{FR} \rangle$ for $\nu_e$ at $t=10$ ms post bounce in 1D. 
The numbers of energy grid points employed are $N_\epsilon=8$, 16, and 32.
The vertical dashed line indicates the radial position of the shock wave.
\label{fig:Ed_High_Resolution_1D_en}}

\begin{center}
\includegraphics[width=\hsize]{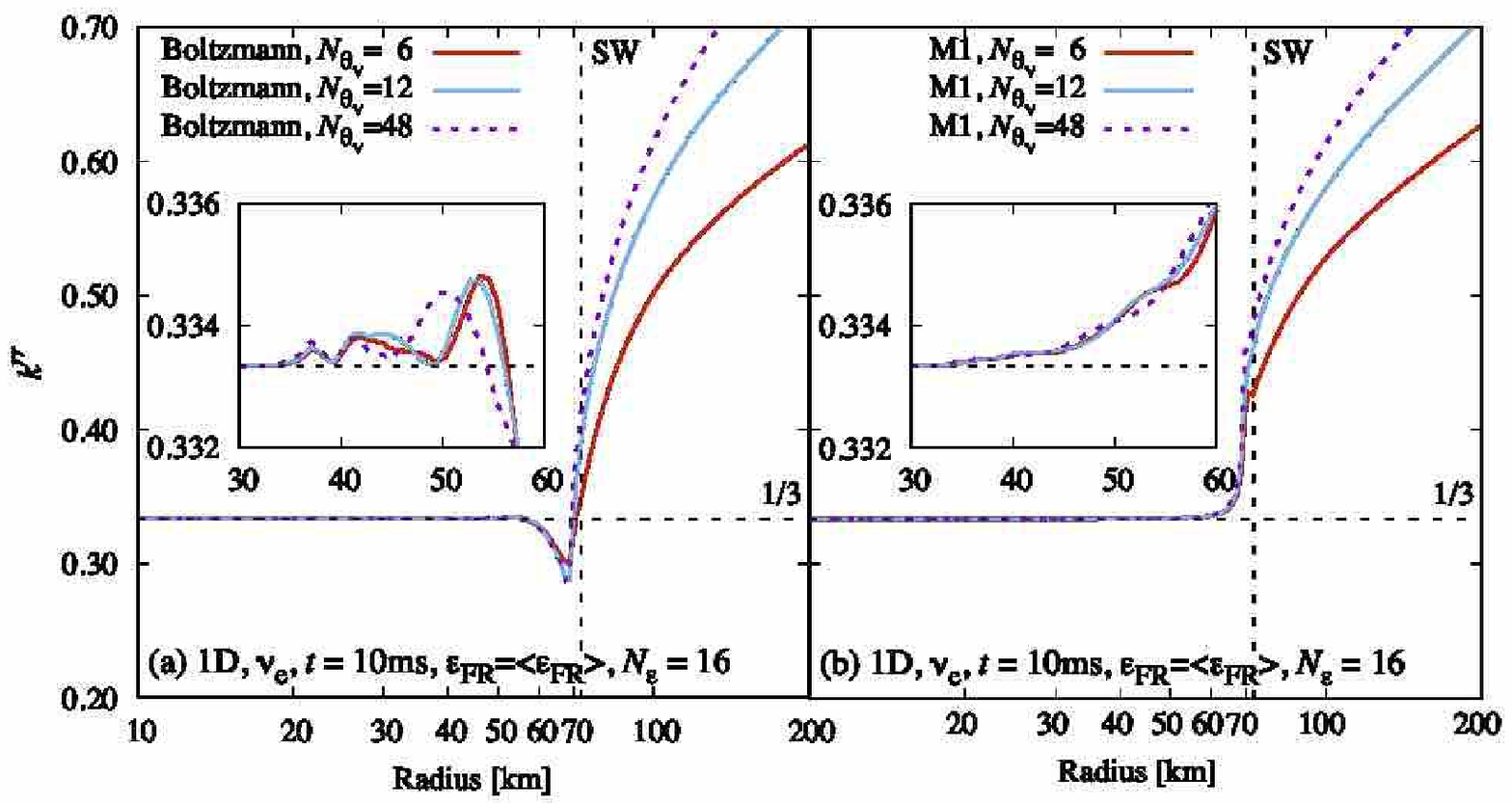}
\end{center}
\caption{The angular-resolution dependence of $k^{rr}$ at $\epsilon_\mathrm{FR}=\langle \epsilon_\mathrm{FR} \rangle$ for $\nu_e$ at $t=10$ ms post bounce in 1D. 
The numbers of the angular grid points employed are for $N_{\theta_\nu}=6$, 12, and 48.
The vertical dashed line indicates the radial position of the shock wave.
\label{fig:Ed_High_Resolution_1D_an}}
\end{figure}

\begin{figure*}[ht!]
\begin{center}
\includegraphics[width=\hsize]{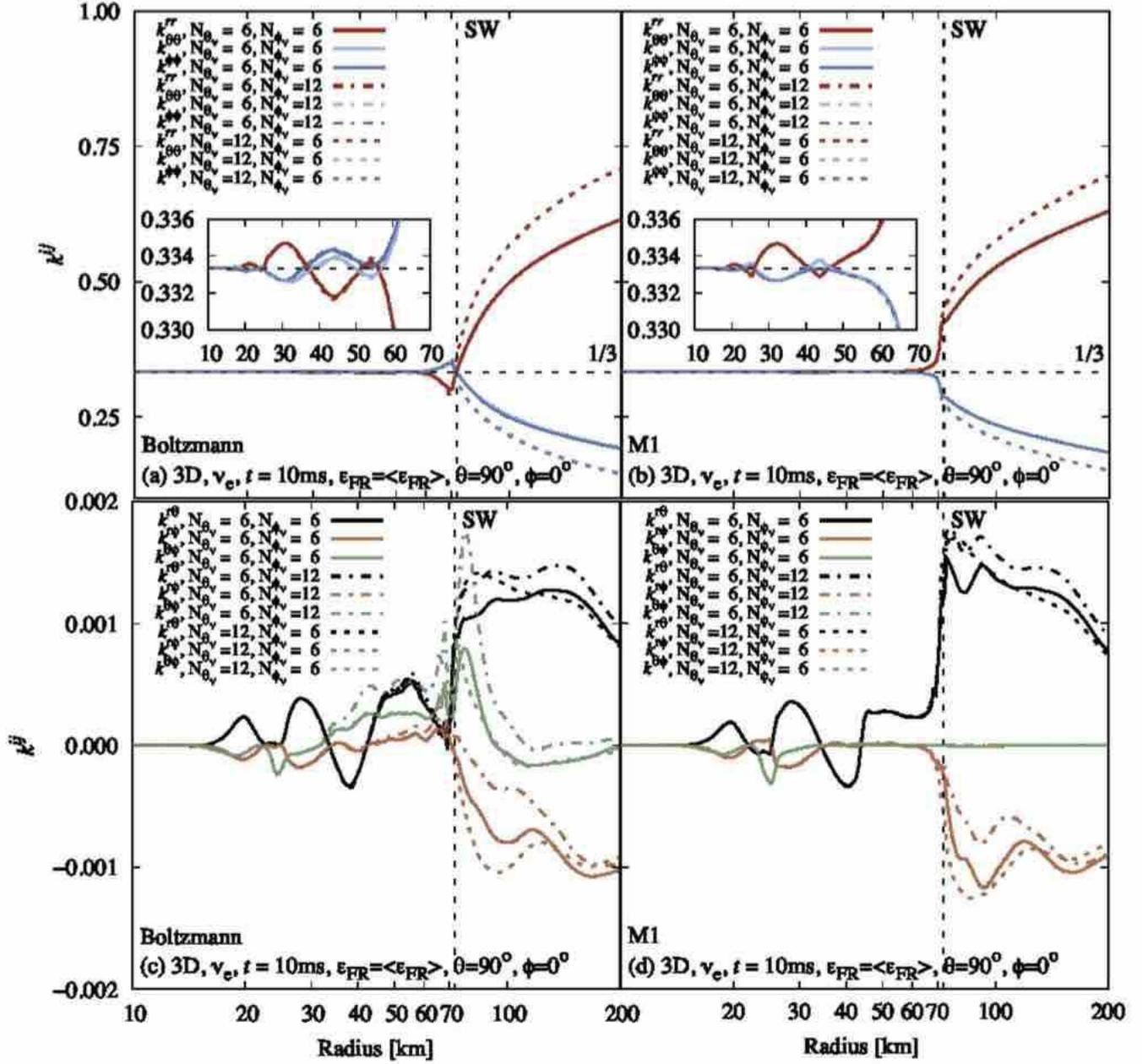}
\end{center}
\caption{
The radial profiles of the diagonal (upper panels) and off-diagonal (lower panels) components of the Eddington tensor $k^{ij}$ (left panels) and $k^{ij}_{M1}$ (right panels) at $\epsilon_\mathrm{FR}=\langle\epsilon_\mathrm{FR}\rangle$ for $\nu_e$ at $t=10$ ms post bounce in 3D.
The numbers of the angular grid points employed are $(N_{\theta_\nu}, N_{\phi_\nu})=(6,6)$, $(6,12)$, and $(12,6)$, denoted in solid, dashed-dotted, and dashed lines, respectively.
The vertical dashed line indicates the radial position of the shock wave.
\label{fig:Ed_High_Resolution}}
\end{figure*}

\begin{figure*}[ht!]
\begin{center}
\includegraphics[width=\hsize]{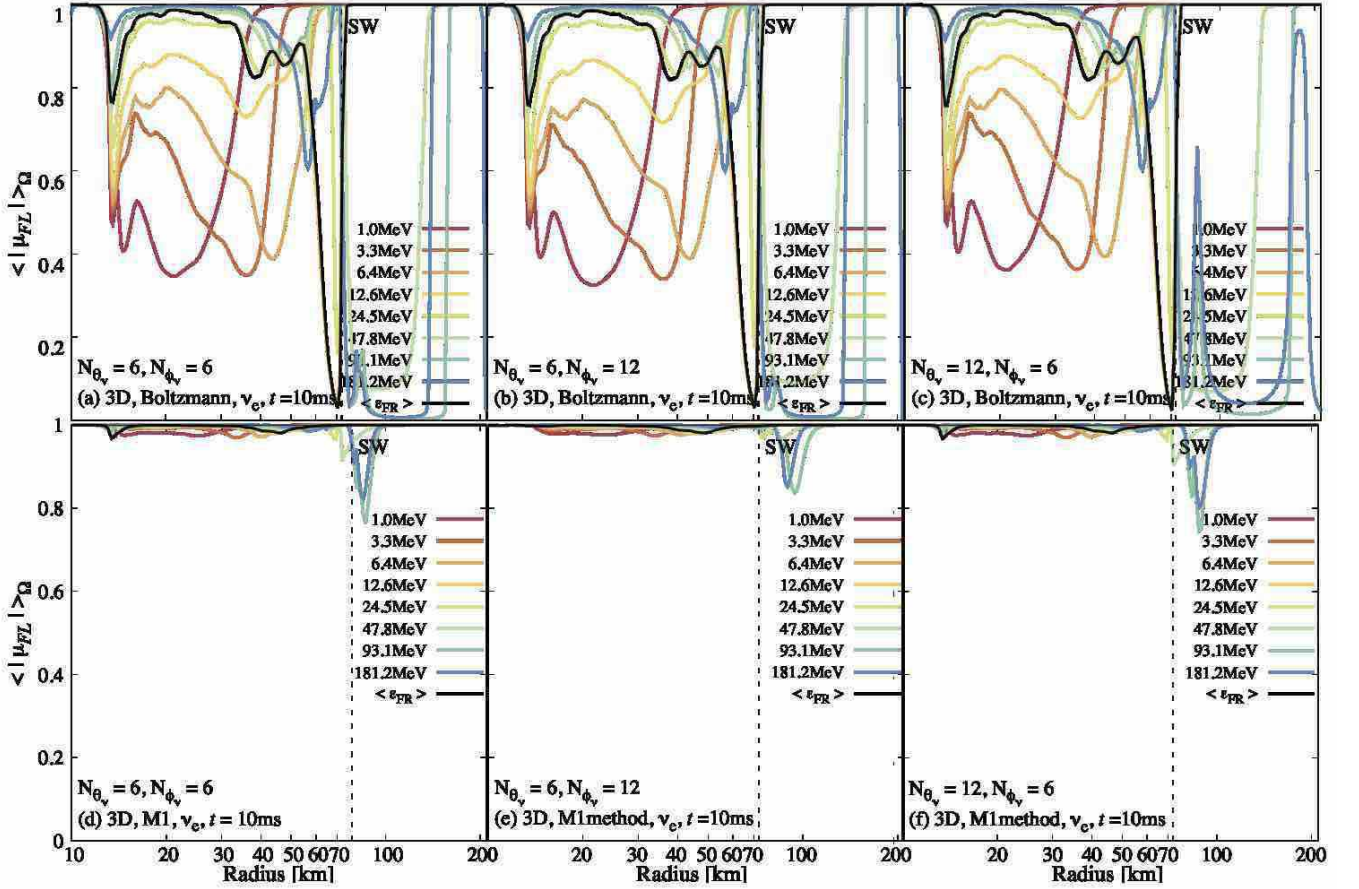}
\end{center}
\caption{The radial profiles of $\langle|\mu_{FL}|\rangle_\Omega$ for $\nu_e$ at $t=10$ ms post bounce for (a) $(N_{\theta_\nu}, N_{\phi_\nu})=(6, 6)$, (b) $(N_{\theta_\nu}, N_{\phi_\nu})=(6, 12)$, and (c) $(N_{\theta_\nu}, N_{\phi_\nu})=(12, 6)$, where $\langle|\mu_{FL}|\rangle_\Omega$ is the absolute value of $\mu_{FL}$, the cosine of the angle between $F$ and $L$, averaged over the whole solid angle $\Omega$.
The vertical dashed line indicates the radial position of the shock wave.
\label{fig:In_High_Resolution}}
\end{figure*}

\begin{figure*}[ht!]
\begin{center}
\includegraphics[width=\hsize]{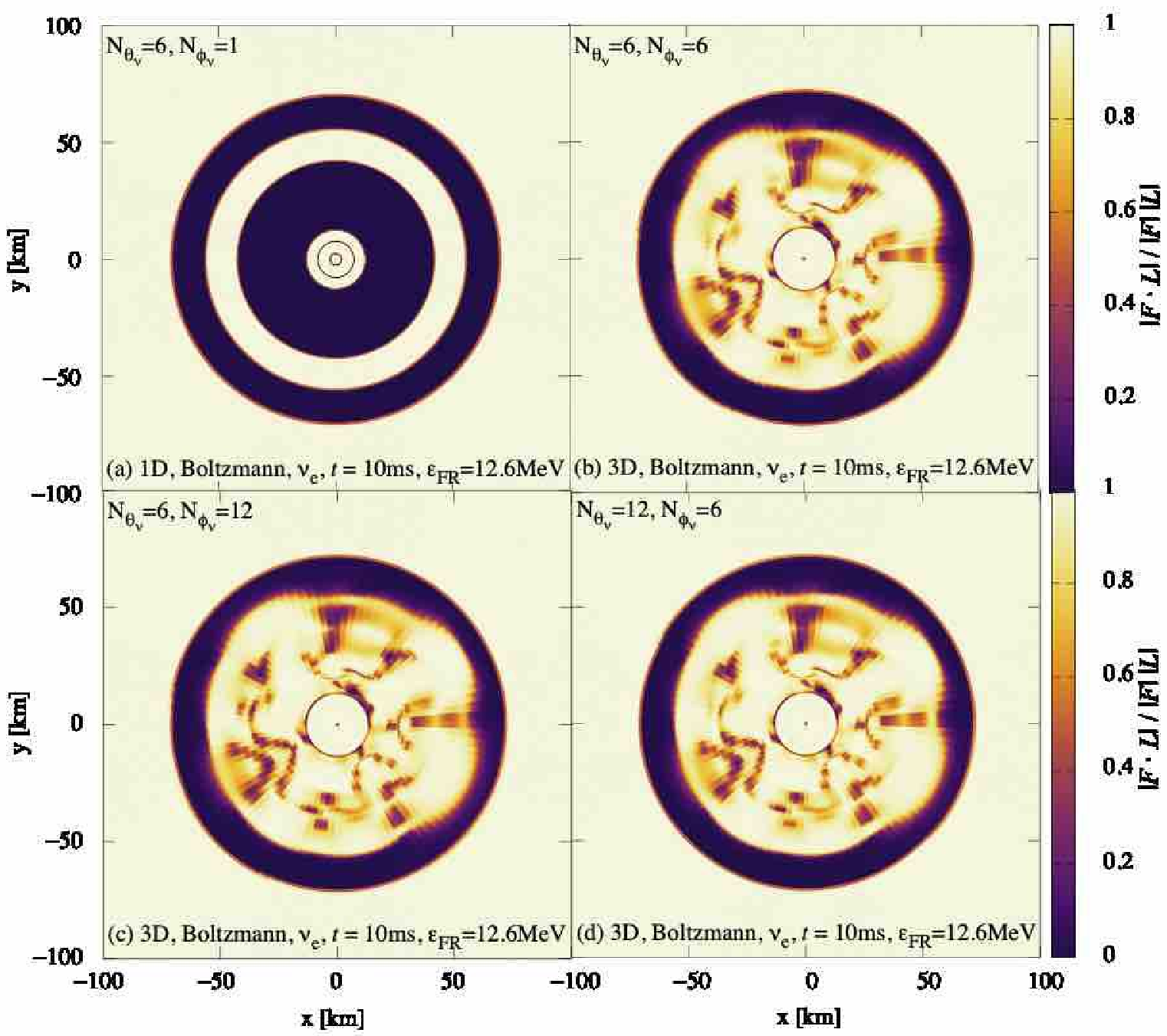}
\end{center}
\caption{The color maps of $|\mu_{FL}|$ at $\epsilon_\mathrm{FR}=12.6$ MeV for $\nu_e$ at $t=10$ ms post bounce for (b) $(N_{\theta_\nu}, N_{\phi_\nu})=(6, 6)$, (c) $(N_{\theta_\nu}, N_{\phi_\nu})=(6, 12)$, and (d) $(N_{\theta_\nu}, N_{\phi_\nu})=(12, 6)$.
For reference the 1D result is also shown in (a).
\label{fig:In_High_Resolution_2D_en007}}
\end{figure*}

\begin{figure*}[ht!]
\begin{center}
\includegraphics[width=\hsize]{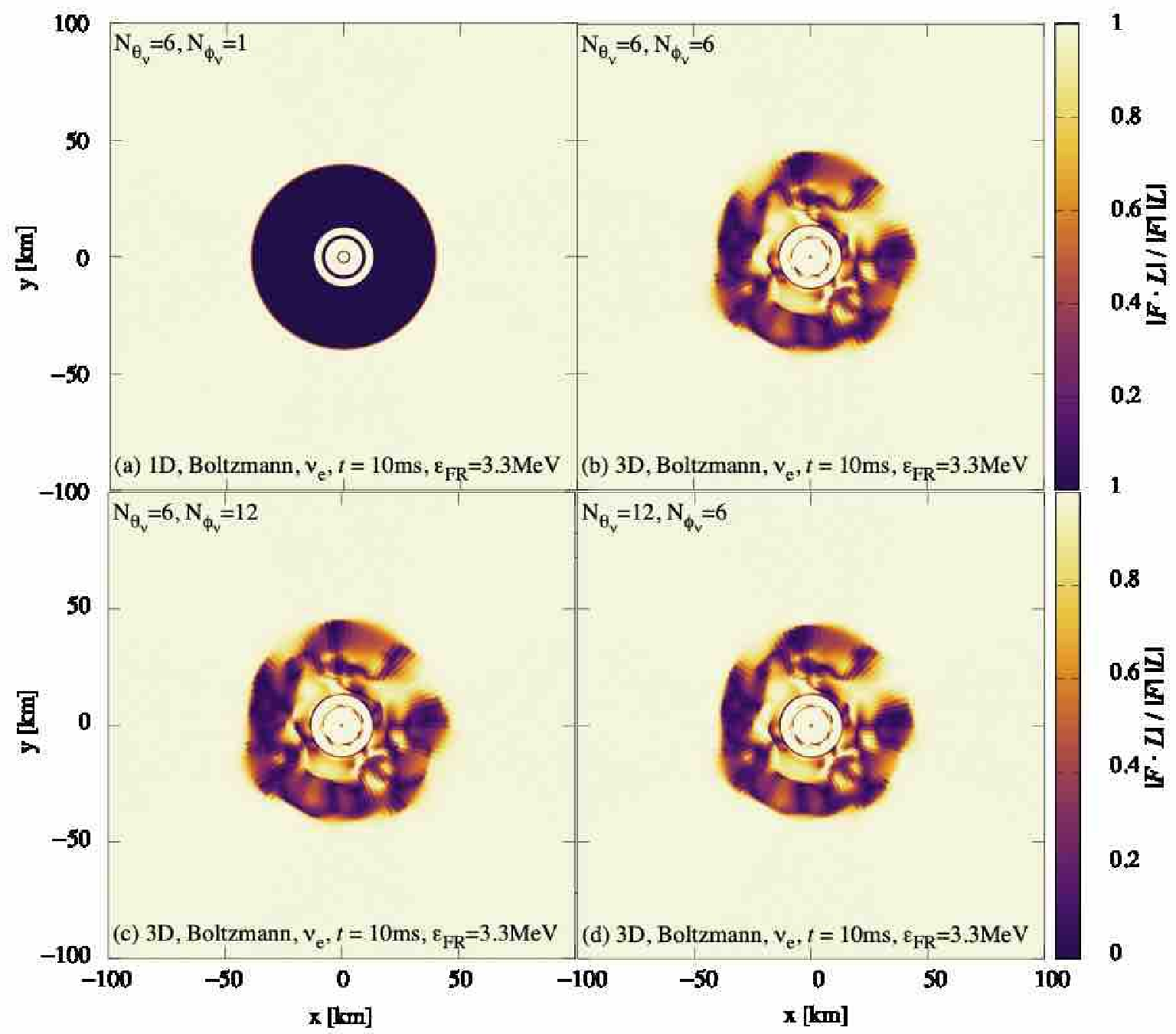}
\end{center}
\caption{The color maps of $|\mu_{FL}|$ at $\epsilon_\mathrm{FR}=3.3$ MeV for $\nu_e$ at $t=10$ ms post bounce for (b) $(N_{\theta_\nu}, N_{\phi_\nu})=(6, 6)$, (c) $(N_{\theta_\nu}, N_{\phi_\nu})=(6, 12)$, and (d) $(N_{\theta_\nu}, N_{\phi_\nu})=(12, 6)$.
For reference the 1D result is also shown in (a).
\label{fig:In_High_Resolution_2D_en003}}
\end{figure*}

In this appendix, we report some of the results of resolution tests we have conducted that we think are of relevance for this paper. We refer readers also to our earlier publications for more tests \citep{Richers2017, Nagakura2018, Harada2019}.

The canonical number of numerical grid points deployed in this paper is $N_r \times N_\theta \times N_\phi \times N_\epsilon \times N_{\theta_\nu} \times N_{\phi_\nu} = 256 \times 48 \times 96 \times 16 \times 6 \times 6$, where $N_r$, $N_\theta$, $N_\phi$, $N_\epsilon$, $N_{\theta_\nu}$, and $N_{\phi_\nu}$ are the grid number of radial, polar and azimuthal angles in space, neutrino energy, polar and azimuthal angles in momentum space, respectively.
Note that the grid spacing of the radial mesh is the same as in \citet{Nagakura2018}, while the computational domain is confined to $r \le 200$ km; the number of numerical grid points in momentum space is reduced from $N_\epsilon \times N_{\theta_\nu} \times N_{\phi_\nu}=20\times10\times6$ in \citet{Nagakura2018} to $16\times6\times6$ in this paper.

We first conduct the resolution test for the spatial angular mesh in 2D.
Comparing the results among $N_\theta=48$, 64, and 96, we find that at $t=10$ ms the prompt convection grows with three coherent vortex rings for $N_\theta=48$ whereas the number increases to four for $N_\theta=64$.
At $N_\theta=96$ we observe three highly deformed and less coherent rings.
We also observe that modes with smaller sizes grow more rapidly, reaching the nonlinear phase earlier, as the spatial resolution increases.
The differences are not so large to affect the conclusions in this paper, though.
We hence set $N_\theta=48$ throughout the paper. 
The number of the azimuthal grid points $N_\phi$ is chosen to be twice $N_\theta$.

Next we perform 1D simulations to examine effects of the energy and angular resolutions in momentum space.
\citet{Richers2017} conducted similar resolution tests, comparing the results from the Boltzmann simulations with those obtained with their Monte Carlo code.
They employed three different grids: $(N_\epsilon, N_{\theta_\nu})$ = (20, 10), (40, 20), and (80, 40).
Their results indicate that even $(N_\epsilon, N_{\theta_\nu})=(80, 40)$ may not be large enough in the discrete ordinates code (Fig.~8 in \citet{Richers2017}).
Since they changed both $N_\epsilon$ and $N_{\theta_\nu}$ at the same time, however, it is not clear which resolution is more important for their results.
Here we vary each of the two numbers individually.
Figure \ref{fig:Ed_High_Resolution_1D_en} shows the energy-resolution dependence of (a) $k^{rr}$ and (b) $k^{rr}_{M1}$, in which $N_\epsilon=8$, 16, and 32.
The radial profiles of $k^{rr}$ for $N_\epsilon=16$ are in good agreement with those for $N_\epsilon=32$, with deviations of the order of $O(10^{-3})$.
This is also the case for $k^{rr}_{M1}$.
We hence conclude that $N_\epsilon=16$ is good enough for our analysis in this paper.
Note also that this is still larger than the number of energy grid points in most supernova simulations.

On the other hand, Figure \ref{fig:Ed_High_Resolution_1D_an} shows the angular resolution dependence of (a) $k^{rr}$ and (b) $k^{rr}_{M1}$, in which we deploy $N_{\theta_\nu}=6$, 12, and 48.
In the optically thin region for $\epsilon_{FR}=\langle\epsilon_{FR}\rangle$ at $r > 70$ km, $k^{rr}$ becomes larger with increasing $N_{\theta_\nu}$; it is obvious that $N_{\theta_\nu} = 6$ is not large enough; it is interesting that $k^{rr}_{M1}$ is smaller than $k^{rr}$ for $N_{\theta_\nu}=12$ and 48 while it is larger for $N_{\theta_\nu}=6$.
It should be noted, however, that the Eddington tensor evaluated with the M1 closure approximation is based on the same numerical data obtained by the Boltzmann simulation. 
Deeper inside at $r= 30\sim60$ km, the radial profiles of $k^{rr}$ and $k^{rr}_{M1}$ for different values of $N_{\theta_\nu}$ both indicate again that $N_{\theta_\nu} = 6$ is not large enough to get numerical convergence.
However, the important feature that there exists a region at $r\sim55-70$ km, in which $k^{rr} < 1/3$ while $k^{rr}_{M1} > 1/3$ is common to the three cases with $N_{\theta_\nu}=6$, 12, and 48. 
We hence consider that $N_{\theta_\nu}=6$ is good enough for the principal-axes analysis of the Eddington tensor in this paper.

Now we turn to the 3D case and consider the angular resolution in momentum space.
The higher-resolution runs with $(N_{\theta_\nu}, N_{\phi_\nu})=(6, 12)$ and $(12,6)$ are initiated from $t=9.5$ ms post bounce, interpolating the result for the original resolution with $(N_{\theta_\nu}, N_{\phi_\nu})=(6,6)$.
Figure \ref{fig:Ed_High_Resolution} shows the radial profiles of diagonal (upper panels) and off-diagonal (lower panels) components of the Eddington tensor $k^{ij}$ (left panels) and $k^{ij}_{M1}$ (right panels) at $\epsilon_\mathrm{FR}=\langle\epsilon_{FR}\rangle$ for $\nu_e$ at $t=10$ ms.
We find for the diagonal components that the larger $N_{\theta_\nu}$ raises $k^{rr}$ and lowers $k^{\theta\theta}$ and $k^{\phi\phi}$ at $r > 70$ km, which is essentially the same as what we saw above in the 1D tests, while the higher value of $N_{\phi_\nu}$ does not induce much change for them in the whole region (Fig.~\ref{fig:Ed_High_Resolution} (a)).
As we mentioned in the previous paragraph for the 1D resolution tests, the region at $r>70$ km is optically thin for $\epsilon_{FR}=\langle\epsilon_{FR}\rangle$ and the ellipsoids are elongated vertically.
Those are the features better captured with higher resolutions.
At smaller radii $r<70$ km the value of $N_{\theta_\nu}$
less affects the diagonal components of the Eddington tensor just as in the 1D resolution tests. 
In fact the important feature that the horizontally wide ellipsoids occur is unaffected by the resolution. 
This is also the case for $k^{ij}_{M1}$ in the sense that it produces almost always vertically long ones irrespective of the angular resolutions.
The off-diagonal components, on the other hand, are also affected by the angular resolutions, particularly at $r > 70$ km for the same reason (Fig.~\ref{fig:Ed_High_Resolution} (c)).
Note that they are much smaller than the diagonal components (see the difference in the scales of the vertical axes in the left panels). It is interesting that $k^{\theta\phi}$ is somewhat more sensitive to the resolution in $\phi_\nu$ direction.
We also point out that it is insensitive to $N_{\theta_\nu}$ and $N_{\phi_\nu}$ that $k^{rr}_{M1}$ ($k^{\theta\theta}_{M1}$ and $k^{\phi\phi}_{M1}$) increases (decrease) monotonically just behind the shock wave (Fig.~\ref{fig:Ed_High_Resolution} (b)) and that $k^{\theta\phi}_{M1}$ is much smaller than $k^{r\theta}_{M1}$ and $k^{r\phi}_{M1}$ at $r > 30$ km (Fig.~\ref{fig:Ed_High_Resolution} (d)).

Now we investigate how the resolution dependence observed in Figure~\ref{fig:Ed_High_Resolution} is reflected in the principal-axes analysis for the Eddington tensor.
Figure~\ref{fig:In_High_Resolution} shows the radial profiles of $\langle|\mu_{FL}|\rangle_\Omega$ for $k^{ij}$ and $k^{ij}_{M1}$ at various energies. 
It is recalled that $|\mu_{FL}| = 0$ and 1 mean respectively that the longest principal axis of the Eddington tensor is perpendicular and parallel to the energy flux.
We find that the results for the higher resolutions agree well with that for the original resolution below the shock wave at every energy  (Fig.~\ref{fig:In_High_Resolution} (a), (b) and (c)).
The results for the M1 closure approximation share the same feature of $\langle|\mu_{FL}|\rangle_\Omega \sim 1$ among the different resolutions (Fig.~\ref{fig:In_High_Resolution} (d), (e) and (f)).
Figures~\ref{fig:In_High_Resolution_2D_en007} and \ref{fig:In_High_Resolution_2D_en003} present the color maps of $|\mu_{FL}|$ at $\epsilon_\mathrm{FR}=12.6$ and 3.3 MeV, respectively, on the equatorial plane for $\nu_e$ at $t=10$ ms in the Boltzmann simulations.
The 1D results are also shown for reference.
The thick colors indicate the regions where the longest principal axis of the ellipsoid is perpendicular to the energy flux.
In the 1D case for $\epsilon_\mathrm{FR}=12.6$ MeV, there are two such regions (Fig.~\ref{fig:In_High_Resolution_2D_en007} (a)).
In the inner region at $\Lambda \sim 0.01$ the ellipsoids are almost spherical with aspect ratios of $\lesssim 1.004$, while in the outer region at $\Lambda \sim 0.1$ they have higher aspect ratios of $\lesssim 1.147$.
In the 1D simulation for $\epsilon_\mathrm{FR}=3.3$ MeV, on the other hand, there exists one thick colored region at $0.1 \lesssim \Lambda \lesssim 1$ (Fig.~\ref{fig:In_High_Resolution_2D_en003} (a)), where the ellipsoids have aspect ratios of $\lesssim 1.042$.
The appearance patterns of the thick colored regions in the 3D runs agree well among all the resolutions: $(N_{\theta_\nu}, N_{\phi_\nu}) = (6,6), (12,6)$, and (6,12) for both $\epsilon_\mathrm{FR}=12.6$ (Fig.~\ref{fig:In_High_Resolution_2D_en007} (b), (c), and (d)) and 3.3 MeV (Fig.~\ref{fig:In_High_Resolution_2D_en003} (b), (c), and (d)).
The outer circular region is what we found below the shock front commonly irrespective of the dimension of the simulations. 
The inner zone observed in the 1D case was shredded into smaller pieces by the convective motions of matter in 3D. 
The important thing, as we mentioned, is that the pattern is essentially the same among the different resolutions.
We focus in this paper only on these features that are rather insensitive to the grid resolution.
Note that the Eddington tensor corresponds to the second angular moments of the distribution function, that is, rather low order, and its ellipsoid in particular reflects not very fine but global features in the angular distribution in momentum space. 
It is hence not surprising that even the angular grid with $(N_{\theta_\nu}, N_{\phi_\nu})=(6,6)$ can capture those low-order features rather well except in the optically thin region.
It is true that we have not yet seen the numerical convergence as observed in Figure~\ref{fig:Ed_High_Resolution} but we believe that the remaining errors do not change the conclusions in this paper.
However, they are based on the single 3D simulation and we certainly need to confirm them with more simulations for different post-bounce times, other progenitor models, and different spatial and momentum resolutions in the future.

\section{Eddington Tensor \label{sec:eddington}}

In this section, we give some definitions of the angular moments of the distribution function in phase space as well as of the Eddington tensor, following the moment formalism described in \citet{Thorne1981} and \citet{Shibata2011}.
An unprojected second moment is first defined in an arbitrary frame as
\begin{eqnarray}
M^{\alpha\beta}(\epsilon_\mathrm{FR}) \equiv \int f(p^\alpha)\delta\left(\frac{\epsilon_\mathrm{FR}^3}{3}-\frac{(-u_\alpha p^\alpha)^3}{3}\right)p^\alpha p^\beta dV_p,
\label{eq:moment}
\end{eqnarray}
where $f$, $p^\alpha$, $u^\alpha$, and $dV_p$ are the neutrino distribution function, four-momentum of neutrinos, four-velocity of medium, and invariant integration element, respectively.
Note that $\epsilon_\mathrm{FR}$ is a parameter to be specified, for example, as the neutrino energy measured in the fluid-rest frame as adopted here.
The Greek indices $\alpha$, $\beta$, and $\gamma$ run over 0, 1, 2, 3, representing the time and space components.
The second angular moment can be written as
\begin{equation}
M^{\alpha\beta}(\epsilon_\mathrm{FR}) = J(\epsilon_\mathrm{FR})u^{\alpha}u^{\beta}+H^\alpha(\epsilon_\mathrm{FR}) u^\beta + H^\beta(\epsilon_\mathrm{FR}) u^\alpha + L^{\alpha\beta}(\epsilon_\mathrm{FR}),\label{eq:moment2}
\end{equation}
where the energy density $J$, energy flux $H$, and radiation pressure tensor $L$ are the variables projected on to the fluid-rest flame.
In this study, in which general relativity is ignored, they are calculated as follows
\begin{align}
& J(\epsilon_\mathrm{FR})=J_\mathrm{FR}(\epsilon_\mathrm{FR}),\\
&H^\alpha(\epsilon_\mathrm{FR})=\Lambda^{\alpha}_{\beta}H^\beta_\mathrm{FR}(\epsilon_\mathrm{FR}),\\
&L^{\alpha\beta}(\epsilon_\mathrm{FR})=\Lambda^\alpha_\gamma \Lambda^\beta_\delta L_\mathrm{FR}^{\gamma\delta}(\epsilon_\mathrm{FR}),
\end{align}
where $\Lambda^\alpha_\beta$ is an appropriate Lorentz transformation; the energy density $J_\mathrm{FR}$, energy flux $H_\mathrm{FR}$, and radiation pressure tensor $L_\mathrm{FR}$ on the right hand side are 
derived from the numerical integration of the distribution function obtained in the Boltzmann simulation.
The same moment can be also expressed as 
\begin{equation}
M^{\alpha\beta}(\epsilon_\mathrm{FR}) =
E(\epsilon_\mathrm{FR})n^{\alpha}n^{\beta}+F^\alpha(\epsilon_\mathrm{FR}) n^\beta + F^\beta(\epsilon_\mathrm{FR}) n^\alpha + P^{\alpha\beta}(\epsilon_\mathrm{FR}),\label{eq:moment3}
\end{equation}
where $n^\alpha$ is a unit timelike vector orthogonal to the hypersurface of constant coordinate time.
The energy density $E$, energy flux $F$, and radiation pressure tensor $P$ on the right hand side of this equation are the variables projected on to the laboratory frame.
The Eddington tensor $k^{ij}$ is then defined as
\begin{eqnarray}
k^{ij}(\epsilon_\mathrm{FR})=\frac{P^{ij}(\epsilon_\mathrm{FR})}{E(\epsilon_\mathrm{FR})}, \label{eq:kij}
\end{eqnarray}
where the index $i$ and $j$ run over 1, 2, 3, corresponding to the space components.

In the truncated moment scheme, the time evolution equations of $E$ and $F$ are normally solved with the algebraic closure relation imposed by hand.
In this paper, we investigate the M1 closure relation, the most favorite choise these days, for which the radiation pressure tensor is expressed as
\begin{equation}
P^{ij}_{\rm M1} (\epsilon_\mathrm{FR}) = \frac{3\zeta(\epsilon_\mathrm{FR}) - 1}{2} P^{ij}_{\rm thin} (\epsilon_\mathrm{FR}) + \frac{3(1-\zeta(\epsilon_\mathrm{FR}))}{2} P^{ij}_{\rm thick} (\epsilon_\mathrm{FR}),
\label{eq:Pijm1}
\end{equation}
where $\zeta$ is the variable Eddington factor given, for example \citep{Levermore1984}, as
\begin{eqnarray}
\zeta(\epsilon_\mathrm{FR}) = \frac{3+4\bar{F}(\epsilon_\mathrm{FR})^2}{5+2\sqrt{4-3\bar{F}(\epsilon_\mathrm{FR})}},
\label{eq:zeta}
\end{eqnarray}
where the flux factor is defined, following \citet{Shibata2011}, 
\begin{eqnarray}
\bar{F}(\epsilon_\mathrm{FR}) = \sqrt{\frac{h_{\alpha\beta} H^\alpha(\epsilon_\mathrm{FR}) H^\beta(\epsilon_\mathrm{FR})}{J(\epsilon_\mathrm{FR})^2}}.
\end{eqnarray}
The optically thin limit of $P^{ij}$ is
\begin{equation}
P^{ij}_{\rm thin}(\epsilon_\mathrm{FR}) = E(\epsilon_\mathrm{FR}) \frac{F^i(\epsilon_\mathrm{FR})F^j(\epsilon_\mathrm{FR})}{F(\epsilon_\mathrm{FR})^2}, \label{eq:Pthin}
\end{equation}
whereas the optically thick limit is
\begin{eqnarray}
P^{ij}_{\rm thick} (\epsilon_\mathrm{FR}) = J(\epsilon_\mathrm{FR}) \frac{\gamma^{ij}+ 4 V^i V^j}{3} + H^i(\epsilon_\mathrm{FR}) V^j + V^i H^j(\epsilon_\mathrm{FR}), \label{eq:Pthick}
\end{eqnarray}
where $V^i$ and  $\gamma^{ij}=g^{ij}+n^i n^j$ stand for the three-dimensional fluid velocity and the spatial metric on the hypersurface of constant coordinate time, respectively.
The Eddington tensor in the M1 prescription is defined as
\begin{eqnarray}
k^{ij}_{M1} (\epsilon_\mathrm{FR})=\frac{P^{ij}_{M1} (\epsilon_\mathrm{FR})}{E(\epsilon_\mathrm{FR})}. \label{eq:kijM1}
\end{eqnarray}

\section{Jacobi method \label{sec:jacobi}}
Jacobi's approach is one of the methods to obtain eigenvalues and eigenvectors of real symmetric matrices \citep[e.g.][]{Vetterling1992}.
Let us consider to diagonalize a real symmetric matrix $\mathbf{A}$.
If the maximum absolute value of the off-diagonal components of $\mathbf{A}$ is given by the $pq$ component, the matrix $\mathbf{A}$ is transformed to
\begin{eqnarray}
\mathbf{A}'= \mathbf{P}_{pq}^t\cdot \mathbf{A}\cdot \mathbf{P}_{pq},
\label{eq:UAU}
\end{eqnarray}
where $\mathbf{P}_{pq}$ is the basic Jacobi rotation matrix, 
\begin{equation}
\mathbf{P}_{pq}=
\left(
\begin{array}{ccccccc}
1      & \cdots &      0     & \cdots & 0          & \cdots & 0      \\
\vdots & \ddots & \vdots     &        & \vdots     &        & \vdots \\
0      & \cdots & \cos\theta & \cdots & \sin\theta & \cdots & 0      \\
\vdots &        & \vdots     & \ddots & \vdots     &        & \vdots \\
0      & \cdots &-\sin\theta & \cdots & \cos\theta & \cdots & 0      \\
\vdots &        & \vdots     &        & \vdots     & \ddots & \vdots \\
0      & \cdots &      0     & \cdots & 0          & \cdots & 1      \\
\end{array}
\right).
\label{eq:Ppq}
\end{equation}
Note that only the components on the $p$-th and $q$-th rows and columns are transformed as follows:
\begin{equation}
\mathbf{A'}=
\left(
\begin{array}{ccccccc}
       & \cdots & a'_{1p} & \cdots & a'_{1q} & \cdots & 
       \\
\vdots & \ddots & \vdots  &        & \vdots  &        & \vdots \\
a'_{p1}& \cdots & a'_{pp} & \cdots & a'_{pq} & \cdots & a'_{pn}\\
\vdots &        & \vdots  & \ddots & \vdots  &        & \vdots \\
a'_{q1}& \cdots & a'_{qp} & \cdots & a'_{qq} & \cdots & a'_{qn}\\
\vdots &        & \vdots  &        & \vdots  & \ddots & \vdots \\
       & \cdots & a'_{np} & \cdots & a'_{nq} & \cdots &        \\
\end{array}
\right),
\label{eq:Adash}
\end{equation}
\begin{equation}
\begin{array}{l}
a'_{rp} =  a_{rp} \cos\theta - a_{rq} \sin\theta, \ \ \ \  (r\ne p, r\ne q, r = 1, 2, \cdots, n), \\
a'_{rq} =  a_{rq} \cos\theta + a_{rp} \sin\theta, \ \ \ \ (r\ne p, r\ne q, r = 1, 2, \cdots, n), \\
a'_{pp} =  a_{pp} \cos^2\theta + a_{qq} \sin^2\theta - 2 a_{pq}\sin\theta\cos\theta, \\
a'_{qq} =  a_{qq} \cos^2\theta + a_{pp} \sin^2\theta + 2 a_{pq}\sin\theta\cos\theta, \\
a'_{pq} =  a_{pq} (\cos^2\theta  - \sin^2\theta) + (a_{pp}-a_{qq})\sin\theta\cos\theta. \label{eq:apq}\\
\end{array}
\end{equation}
In the Jacobi method, each rotation is chosen so that the off-diagonal $pq$ component should be zero, $a'_{pq}=0$.
Then we have the following relations,
\begin{equation}
\begin{array}{l}
\cos \theta = \sqrt{\frac{1+\alpha}{2}}, \\
\sin \theta = \sqrt{\frac{1-\alpha}{2}}\mathrm{sgn}(\beta\gamma),\\
\alpha=\frac{|\beta|}{\sqrt{\beta^2+\gamma^2}}, \ \ \ \beta = \frac{1}{2}(a_{pp}-a_{qq}), \ \ \ \gamma = -a_{pq}.
\end{array}
\end{equation}
Repeating this operation until all the off-diagonal components vanish (actually fall below $1.0\times 10^{-8}$),
we eventually obtain the diagonal matrix $\mathbf{D}$ that satisfies
\begin{eqnarray}
\mathbf{D}= \mathbf{V}^t\cdot \mathbf{A}\cdot \mathbf{V},
\label{eq:D}
\end{eqnarray}
where $\mathbf{V}$ is the rotation matrix given as
\begin{eqnarray}
\mathbf{V}= \mathbf{P}_1 \cdot \mathbf{P}_2\cdot \mathbf{P}_3 \cdots, 
\label{eq:V}
\end{eqnarray}
with $\mathbf{P}_i$ being one of the successive Jacobi rotation matrices.
The diagonal components of $\mathbf{D}$ finally give the eigenvalues of the original matrix $\mathbf{A}$, and the columns of $\mathbf{V}$ are the corresponding eigenvectors.

\begin{figure*}[b]
\begin{center}
\includegraphics[width=\hsize]{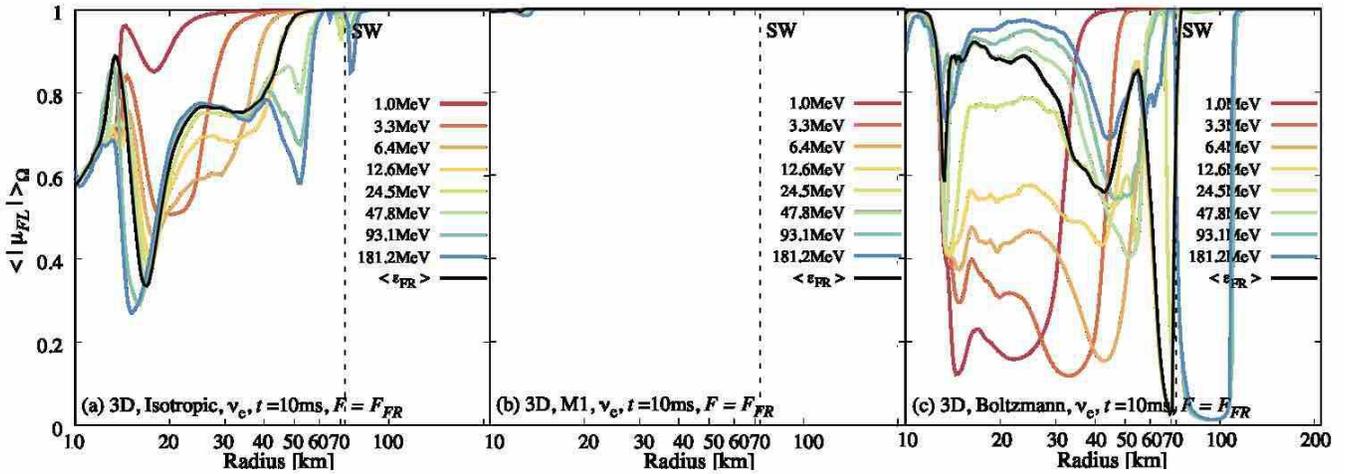}
\end{center}
\caption{The radial profiles of $\langle|\mu_{FL}|\rangle_\Omega$ in the fluid-rest frame for $\nu_e$ at $t=10$ ms post bounce.
(a) $\mu_\mathrm{FL}$ is actually the cosine of the angle between $F$ and $r$, (b) $\mu_\mathrm{FL}$ in the M1 closure approximation, and (c)  $\mu_\mathrm{FL}$ obtained in the Boltzmann simulation.
The vertical dashed line indicates the radial position of the shock wave.
\label{fig:In_FR_10ms}}
\end{figure*}

\section{Principal-Axes Analysis in the Fluid-Rest Frame \label{sec:analysisA}}
In this paper, we analyze the Eddington tensor defined in the laboratory frame (Eq.~\ref{eq:kij}). We can do the same in the fluid rest frame, though. It is in fact easier in the optically thick regime.
Figure~\ref{fig:In_FR_10ms} shows the radial profiles of $\langle|\mu_{FL}|\rangle_\Omega$ in the fluid-rest frame for $\nu_e$ at $t=10$ ms post bounce, where $\langle|\mu_{FL}|\rangle_\Omega$ is the absolute value of $\mu_{FL}$, the cosine of the angle between $F$ and $L$, averaged over the whole solid angle $\Omega$.
It should be mentioned first that if the angular distribution is completely isotropic, the Eddington tensor is proportional to the unit tensor and the longest principal axis is arbitrary but set to be radial in our code; then $\mu_{FL}$ becomes the cosine of the angle between $F$ and $r$.
This never happens in the supernova core in reality, since neutrinos diffuse out of fluid elements to the adjacent ones according to the gradient of energy density even in the optically thick region.
This is demonstrated in Fig.~\ref{fig:In_FR_10ms} (a), in which we show the cosine of the angle between $F$ and $r$ averaged over the solid angle to be compared with the $\langle|\mu_\mathrm{FL}|\rangle_\Omega$ presented in Fig.~\ref{fig:In_FR_10ms} (c) in the same figure.
In the fluid-rest frame, the pressure tensor in the M1 closure approximation has just two terms of $\gamma^{ij}$ and $F^iF^j$, in which case $L$ should be simply parallel to $F$.
This is indicated indeed at $r > 14$ km in Fig.~\ref{fig:In_FR_10ms} (b), in which we show the result for the M1 approximation.
The small deviation from unity at $r < 14$ km is come from $\zeta -1/3 \lesssim O(10^{-8})$, where $\zeta$ is the variable Eddington factor in Eq.~(\ref{eq:zeta}).
Then the first term of Eq.~(\ref{eq:Pijm1}) divided by $E(\epsilon_{FR})$, which determines the direction of $L$ in the fluid-rest frame, becomes lower than $O(10^{-8})$ at $r < 14$ km.
This means that the analysis can be precisely available for $r > 14$ km at $t=10$ ms post bounce since the accuracy of the Jacobi method is $O(10^{-8})$.
It is pointed further that the radial profiles of $\langle|\mu_{FL}|\rangle_\Omega$ in the fluid rest frame for the Boltzmann simulation (Fig.~\ref{fig:In_FR_10ms} (c)) are qualitatively different from those of $\langle|\mu_{FL}|\rangle_\Omega$ in Fig.~\ref{fig:In_FR_10ms} (a) in the transition regions.
This is simply because $L$ is not radial there.
The radius, at which $\langle|\mu_{FL}|\rangle_\Omega$ gets minimum in the fluid-rest frame (Fig.~\ref{fig:In_FR_10ms} (c)), gets larger with the neutrino energy, the feature also observed in the laboratory frame (Fig.~\ref{fig:Inner1D}(a)).
The minimum values in the fluid rest-frame, on the other hand, tend to be lower than those in the laboratory frame.
Regardless of the reference frame, the message here is that there are regions, in which the longest principal axis of the ellipsoid derived from the Eddington tensor is perpendicular to the energy flux in the Boltzmann simulation.

\bibliographystyle{aasjournal}
\bibliography{library}

\end{document}